\documentclass[traditabstract]{aa} 
\usepackage{graphicx}
\usepackage{tikz}
\usepackage{natbib}
\bibpunct{(}{)}{;}{a}{}{,}
\begin{document}
   \title{Gamma-Ray emission from SN2014J near maximum optical light\thanks{ Based on observations with \emph{INTEGRAL}, an ESA project with instruments and the science data center funded by ESA member states (especially the PI countries: Denmark, France, Germany, Italy, Switzerland, and Spain), the Czech Republic, and Poland and with the participation of Russia and USA. }
}


\author{J. Isern\inst{1} 
\and P. Jean \inst{2,3} 
\and E. Bravo \inst{4} 
\and J. Kn\"odlseder\inst{2,3} 
\and F. Lebrun \inst{5}
\and E. Churazov \inst{ 6,7}
\and R. Sunyaev \inst{6,7}
\and A. Domingo \inst{8} 
\and C. Badenes \inst{9}
\and D. H. Hartmann \inst{10}
\and P. Hoeflich \inst{11} 
\and M. Renaud \inst{12}
\and S. Soldi \inst{5}
\and N. Elias--Rosa \inst{13,1} 
\and M. Hernanz \inst{1}
\and I. Dom\'{\i}nguez  \inst{14}
\and D. Garc\'{\i}a-Senz \inst{15} 
\and G.G. Lichti \inst{16}
\and G. Vedrenne \inst{2,3}
\and P. Von Ballmoos \inst{2,3}
}

   \institute{Institut de Ci\`encies de l'Espai (ICE-CSIC/IEEC),
             Campus UAB, 08193 Bellaterra, Barcelona, Spain\\
              \email{hernanz@ieec.cat}; \email{isern@ieec.cat}
\and Universit\'e de Toulouse; UPS-OMP; IRAP; Toulouse, France
\and IRAP, 9 Av colonel Roche, BP44346, 31028 Toulouse Cedex 4, France
       \email{pierre.jean@irap.omp.eu}; \email{jknodlsede@irap.omp.eu}; \email{pvb@irap.amp.eu};
       \email{vedrenne@irap.amp.eu}
\and E.T.S.A.V., Univ. Polit\`{e}cnica de Catalunya, c/Pere Serra 1-15, 08173 Sant Cugat
                          del Valles, Spain\\
                           \email{eduardo.bravo@upc.edu}
\and APC, Univ. Paris Diderot, CNRS/IN2P3, CEA/Irfu, Obs. de Paris, Sorbonne Paris Cit\'e,
        10 rue Alice Domon et Leonie Duquet, F-75205 Paris Cedex 13, France
       \email{lebrun@apc.univ-paris7.fr}; \email{ssoldi@apc.in2p3.fr}
\and Space Research Institute (IKI), Proufsouznaya 84/32, Moscow 117997, Russia
\and Max -Planck-Institut for Astrophysics, 
               Karl-Schwarzschild-Strasse 1, D-85741 Garching , Germany\\
       \email{churazov@mpa-garching.mpg.de}; \email{sunyaev@mpa-garching.mpg.de}
\and Centro de Astrobiolog\'{\i}a (CAB-CSIC/INTA),
           P.O. Box 78, 28691 Villanueva de la Ca\~{n}ada, Madrid, Spain \\
       \email{albert@cab.inta-csic.es}
\and Department of Physics and Astronomy \& Pittsburgh Particle Physics, Astrophysics and
        Cosmology Center (PITT-PACC), University of Pittsburgh, Pittsburgh PA15260, USA \\
        \email{badenes@pitt.edu}
\and Department of Physics and Astronomy, Clemson University, Clemson, SC 29634, USA \\
        \email{hdieter@clemson.edu}
\and Physics Department, Florida State University,
        Tallaharssee, FL32306, USA \\
       \email{pah@aastro.physics.fsu.edu}
\and Laboratoire Univers et Particules de Montpellier (LUPM), UMR 5299, Universit\'e de Montpellier II, F-34095, Montpellier, France \\
        \email{mrenaud@lupm.univ-montp2.fr}
\and INAF - Osservatorio Astronomico di Padua, Vicolo dell'Osservatorio 5, I-35122, Padova, Italy \\
        \email nancy.elias@oapd.inaf.it
\and Universidad de Granada, Cuesta del Hospicio sn, E-18071, Granada, Spain \\
        \email {inma@ugr.es}
\and Dept. Fisica i Enginyeria Nuclear, UPC, Compte d'Ugell 187, 08036
                          Barcelona, Spain\\
                          \email{domingo.garcia@upc.edu}
\and Max -Planck-Institut for Extraterrestrial Physics, 
               Giessenbachstrasse 1, D-85741 Garching , Germany\\
        \email{grl@mpe.mpg.de}
            }

   \date{\today}

\abstract{
The optical light curve  of Type Ia supernovae (SNIa) is powered by thermalized gamma-rays produced by the decay of \element[ ][56]{Ni} and \element[ ][56]{Co}, the main radioactive isotopes synthesized by the thermonuclear explosion of a C/O white dwarf. Gamma-rays escaping the ejecta can be used as a diagnostic tool for studying the characteristics of the explosion. In particular, it is expected that the analysis of the early gamma emission, near the maximum of the optical light curve, could provide information about the distribution of the radioactive elements in the debris. In this paper, the gamma data obtained from SN2014J in M82 by the instruments on board of \emph{INTEGRAL} are analyzed taking special care of the impact that the detailed spectral response has on the measurements of the intensity of the lines. The 158 keV emission of \element[ ][56]{Ni} has been detected in SN2014J at $\sim$ 5$\sigma$ at low energy with both ISGRI and SPI around the maximum of the optical light curve.  After correcting the spectral response of the detector, the fluxes in the lines suggest that, in addition to the bulk of radioactive elements buried in the central layers of the debris, there is a plume of \element[ ][56]{Ni}, with a significance of $\sim 3 \sigma$, moving at high velocity and receding from the observer. The mass of the plume is in the range of $\sim 0.03-0.08$ M$_\odot$. No SNIa explosion model had predicted the mass and geometrical distribution of \element[ ][56]{Ni} suggested here. According to its optical properties, SN2014J looks as a normal SNIa. So it is extremely important to discern if it is also representative in the gamma-ray band. 
}

  
   \maketitle
%

\section{Introduction}
Type Ia supernovae (SNIa) are the outcome of the thermonuclear explosion of a carbon/oxygen 
white dwarf in a close binary system. During this explosion significant amounts of 
radioactive  isotopes are produced, the most abundant being \element[ ][56]{Ni} which 
decays to \element[ ][56]{Co} ( $T_{1/2}=  6.08$  days) and further to \element[ ][56]{Fe} 
($T_{1/2} =  77.24$ days). These radioactive nuclei produce gamma-rays that thermalize in 
the ejecta and are thus ultimately responsible for the power of the luminous supernova. The 
most important line-producing nuclear transitions are due to \element[ ][56]{Ni} (158, 480, 
750 and 812 keV) and \element[ ][56]{Co} (847 and 1238 keV). To these photons one must add 
those produced by the annihilation of positrons either directly or through the formation of
 positronium.  As ejecta expansion proceeds, matter becomes more and more transparent and an 
increasing fraction of gamma rays escapes, thus avoiding thermalization. Therefore, gamma-rays 
escaping the ejecta can be used as a diagnostic tool for studying the structure of the exploding 
star and the characteristics of the explosion \citep{clay69,gome98,hoef98,the14}. In particular, the comparison between early and late spectra is especially illustrative since it is sensitive to the distribution 
of \element[ ][56]{Ni} in the debris \citep{gome98}, and can thus advance our understanding of this crucial cosmological tool \citep{iser11,hill13}. The attempt to detect with INTEGRAL  gamma-ray emission from SN2011fe in M101 failed because of its distance ($\sim 6.4$ Mpc) yielded a too faint flux \citep{iser13}. So far, the signatures of $^{56}$Ni and $^{56}$Co decay were observed in hard X-rays and gamma-rays only from the Type II SN1987A in The Large Magellanic Cloud \citep{suny87,matz88,teeg89}. 

SN2014J was discovered by \citet{foss14} on January 21st in M82 (d = 3.5 $\pm$ 0.3 Mpc). The moment of the explosion of SN2014J was estimated to be on January 14.72 UT 2014 \citep{zhen14} or JD 2456672.22. \emph{INTEGRAL} began on observing this source on January 31st, 16.5 days after the explosion  and ended 18th February, 35.2 days after the explosion. Late time observations, ~50 – 100 days after explosion, were also programmed allowing the detection of the \element[ ][56]{Co} emission lines for the first time \citep{chur14,chur14a}. The firm detection in SN2014J of the gamma-ray emission when the \element[ ][56]{Ni} emission is still important \citep{iser14} offers the opportunity to gain insight on the abundances and distribution of these radioactive isotopes. 

\section{\emph{INTEGRAL} observations}

\begin{table}
\caption{\emph{INTEGRAL} observations schedule (IJD: \emph{INTEGRAL} Julian Day). The moment of the explosion of SN2014J, January 14.72 UT 2014, corresponds to JD 2456672.22 or IJD 5127.75. 
}             
\label{table:1}      
\centering                   
\begin{tabular}{c c c c}       
\hline\hline                 

Orbit  &  IJD start & IJD Stop & Days after explosion \\ 
\hline  
1380	& 5144.298 & 5146.956 & 16.5-19.2 \\
1381	& 5147.289 & 5149.941 & 19.6-22.2 \\
1382	& 5150.280 & 5152.699 & 22.6-25.0 \\
1383	& 5153.925 & 5155.922 & 26.2-28.2 \\
1384 & 5156.262 & 5158.486 & 28.6-30.8 \\
1385	& 5159.254 & 5161.899 & 31.6-34.2 \\
1386	& 5162.248 & 5162.858 & 34.4-35.2 \\
\hline                                   
\end{tabular}
\end{table}

\emph{INTEGRAL} is an ESA scientific mission able to operate in gamma-rays, X-rays and visible light \citep{wink03}. It was launched in October 17th 2002 into a highly eccentric orbit with a period of ~3 days, spending most of this time outside the radiation belts. The results presented here were obtained during orbits 1380-1386 as described in Table \ref{table:1} (proposal number 1170001, public; proposal number 1140011, PI: Isern). Orbit 1387 was devoted to calibration and orbit 1388 was affected by a giant solar flare. 

The instruments on board are: i) The OMC camera,  able to operate in the visible band up to a magnitude 18 \citep{mash03}, it was used to obtain the light curve in the V-band allowing an early estimate of the amount of \element[ ][56]{Ni} necessary to account for the shape of the optical light curve, as well as to predict the intensity of the \element[ ][56]{Co} line at late times, ii) the X-ray monitors JEM-X, that work in the range of 3 to 35 keV \citep{lund03} that were used to constrain the continuum emission of SN2014J in this band, and iii) the two main gamma-instruments, SPI, a cryogenic germanium spectrometer able to operate in the energy range of 18 keV - 10 MeV  \citep{vedr03}, and IBIS/ISGRI, an imager able to operate in the energy range of 15keV to 1 MeV \citep{lebr03,uber03}. Below 300 keV IBIS/ISGRI is more sensitive than SPI, a factor $\sim$ 3 in the region of $\sim$ 150 keV \citep{lebr03,roqu03}, but the sensitivity of both instruments is good enough to allow the comparison of the results in the region of 40 to 200 keV.

\subsection{The OMC data}
\begin{figure}
   \centering
 \includegraphics[width=\hsize]{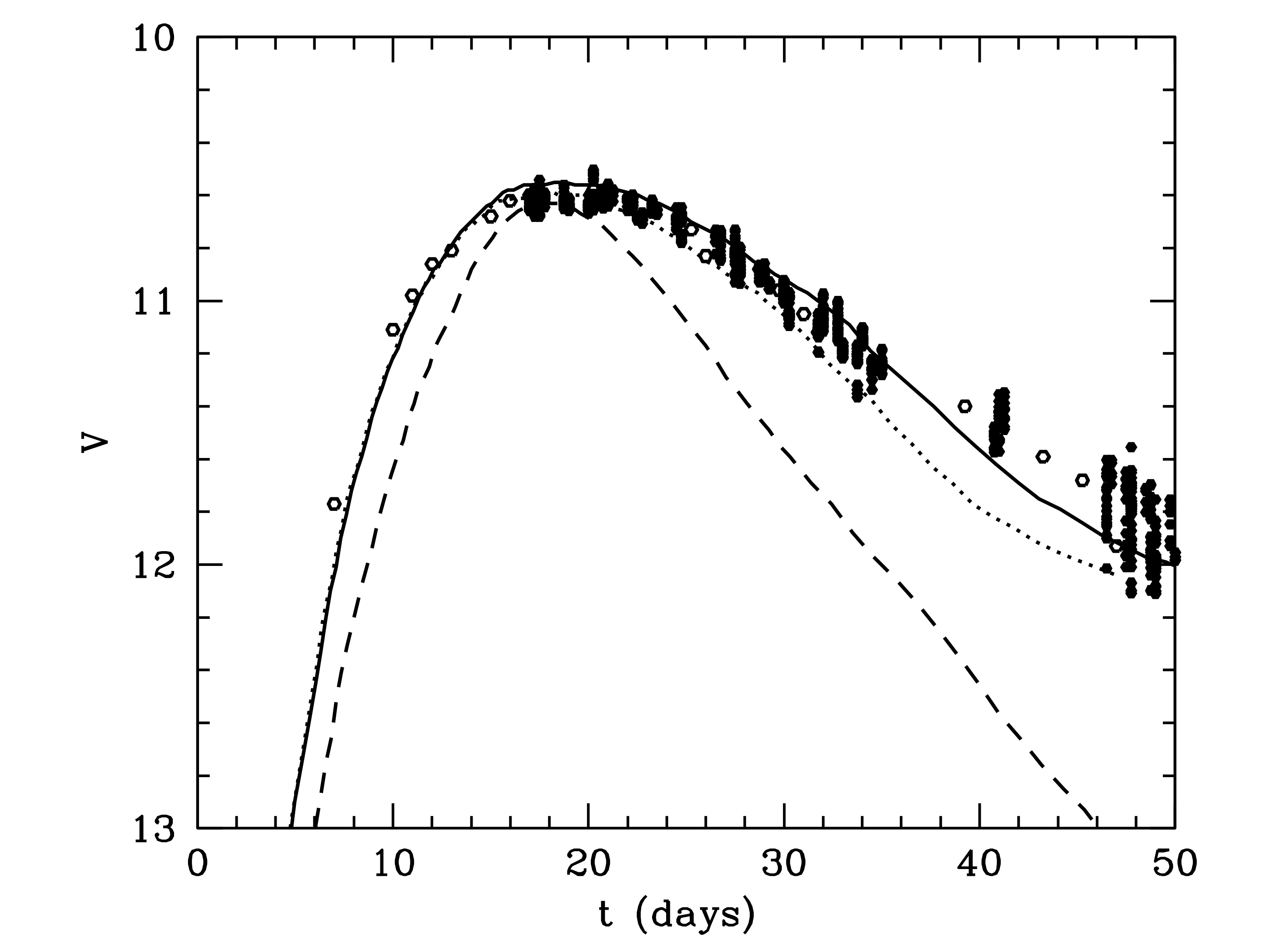}
      \caption{Optical light curve in the V-band versus the time after the explosion. Filled dots represent the data obtained with the OMC camera on board of INTEGRAL without correcting from extinction, empty  dots the data provided by Las Cumbres Observatory Global Telescope Network \citep{mari15}. The solid line is a delayed detonation (DDT) that gives a good fit to the SN2014J OMC light-curve.  It produces 0.65 M$_\odot$ of \element[ ][56]{Ni}, and ejects a total mass of 1.37 M$_\odot$ with a kinetic energy of $1.3 \times 10^{51}$ ergs. The dashed line represents a similar DDT explosion in which a 1.2 M$_\odot$ white dwarf is embedded in a 0.2 M$_\odot$ halo made of carbon and oxygen, and is intended to be representative of merging models (double degenerates, or DD models). The dotted line is for a model fitting the light-curve properties of SN2011fe \citep{iser13}. 
              }
         \label{fig1}
   \end{figure}

Fig.\ref{fig1} displays the light curve obtained with the OMC as well as those provided by several models obtained assuming different parameters and  boundary conditions in order to illustrate the sensitivity of the light curve to different hypothesis about the origin of the supernova. The reduction of the photometric data followed the same procedures as in the case of SN2011fe \citep{iser13}. For SN2014J the main difficulty is the contamination by unresolved stars in M82 as a consequence of the large pixel size of the OMC (17.5"). This problem was overcome by subtracting images of the same region of M82 obtained in 2012.

The peak magnitude of V=10.6 occurred at JD = 2456691.9 $\pm$ 1, 19.7 days after the explosion and, 15 days after maximum, the light curve had dropped by 0.6 mag, in agreement with observations obtained at the Las Cumbres Observatory Global Telescope Network \citep{mari15}. Fig. \ref{fig1} displays the resulting V-band light curve without correcting for extinction. This light curve can be compared to several theoretical models and, in principle, a matching solution can be obtained. Nevertheless, this solution is not unique since similar light curves can be obtained conveniently tuning the different parameters that characterize each model family. Furthermore, environment circumstances, like the presence of circum-stellar material can modify the shape of the light curve as it can be seen in the figure. Just as an example, the model represented by a continuous line in the figure has synthesized $\sim 0.65$ M$_\odot$ of \element[ ][56]{Ni}.

Since the optical and infrared observations of SN2014J \citep{mari15} strongly support the idea of no-mixing in the outer layers, and delayed detonation  (DDT) models\footnote{  Models in which the flame starts at the centre and propagates subsonically making a transition to a supersonic regime when the density is small enough \citep{hoef02}.} predict such a behaviour, a model of this class reasonably fitting the light curve and satisfying the constraints imposed by the late observations of \emph{INTEGRAL} \citep{chur14a} has been selected as a reference. This model, the DDT1P4 model, produces 0.65 M$_\odot$ of \element[ ][56]{Ni} and ejects 1.37 M$_\odot$ of material with a kinetic energy of $1.32 \times 10^{51}$ erg. During the epoch corresponding to $\sim 50-100$ days after the explosion, the 847 and 1238 keV \element[ ][56]{Co} lines obtained with this model exhibit a mean flux of $3.1 \times10^{-4}$ and $2.2 \times 10^{-4}$ cm$^{-2}$s$^{-1}$, are centred at 851 and 1244 keV and have a FWHM of 29 and 42 keV respectively, to be compared with the observed values $(2.34\pm 0.7) \times 10^{-4}$ and $(2.78 \pm 0.7) \times10^{-4}$ cm$^{-2}$s$^{-1}$, $852 \pm 4.5$ and $1255 \pm 7$ keV, and $24 \pm 8$ and $45 \pm 14$ keV respectively \citep{chur14a}. This amount of \element[ ][56]{Ni} is also in agreement with the value, $\sim 0.6$ M$_\odot$ obtained with the mid-infrared observations \citep{tele15}.

\subsection{JEM-X data}
They were analyzed with the same methods that are described in  \citet{iser13}. The flux during revolutions 1380-1386 at the position of the supernova was $1.5 \times10^{-3}$ cm$^{-2}$ s$^{-1}$ in the 3-10 keV band while there was no detection in the 10-25 keV range, with a $3\sigma$ flux upper limit of $6 \times 10^{-4}$ cm$^{-2}$ s$^{-1}$. These fluxes are consistent with the values found in the same position before the explosion  and can be attributed to the combined contribution of compact sources in M82, in particular M82X-1 and X-2 \citep{bach14,sazo14}.

\subsection{SPI data}
The SPI data were cleaned and calibrated with the standard procedure described in section 2.2 of Isern et al. (2013). During this period of observations, science windows showing high rates in the anticoincidence system of SPI when INTEGRAL was exiting the radiation belts were removed ($\sim 2-3$ first science windows per revolution) to avoid systematic errors induced by strong background fluctuations. e.g. see Fig. 5 of \citet{jean03}. 

The behaviour of the instrumental background, produced by the interactions of cosmic-rays and solar protons with the instrument, is very complex, see \citet{jean03} and \citet{weid03} for a detailed discussion. Unfortunately, the two main decay lines of  \element[ ][56]{Ni}, the 158 keV and 812 keV lines, may be affected by two instrumental lines due to decays of \element[ ][47]{Sc} and  \element[ ][58]{Co} that produce lines at 159 keV and 811 keV, respectively, depending on the shift of the  \element[ ][56]{Ni} lines with respect to their canonical energy.

The spatial and temporal modulations produced by the coded mask and dithering allow to reject the background lines as long as their positions and widths are well aligned among the detectors, and the detector pattern (the relative count rate between detectors) is well known. The procedure adopted here consists in adjusting the flux from SN2014J for each energy bin in two steps. 
In the first step, a background count rate was obtained per orbit, detector and energy bin to fix the detector pattern. In the second step, a global background rate factor was fitted per pointing, keeping the detector pattern (i.e. the relative count rate between detectors) fixed to the values determined in the first step. 
 Despite such precautions some residual instrumental lines could remain  and since these lines are intrinsically narrow, any narrow feature of the observed spectrum risks to be confused with them if the background is not correctly modelled.

Indeed, one of the main problems in the interpretation of the data is that the flux extracted in an energy bin $E_{\mathrm{ i}}$ contains not only the source photons emitted at this energy  (later called diagonal terms) but also those emitted with an energy $E > E_{\mathrm{ i}}$ that do not deposit all their energy in the detectors (e.g. Compton edge, backscattering photons - later called off-diagonal terms). This last contribution is not negligible at low energies and can be obtained comparing the extracted spectrum with the theoretical spectra duly convolved with the spectral response of the instrument. 
Therefore, in order to make a meaningful comparison between the spectra measured with SPI and theoretical  models, these ones have to be convolved with the instrumental response to take into account the off-diagonal terms of the spectral response.

   \begin{figure}
   \centering
\includegraphics[width=\hsize]{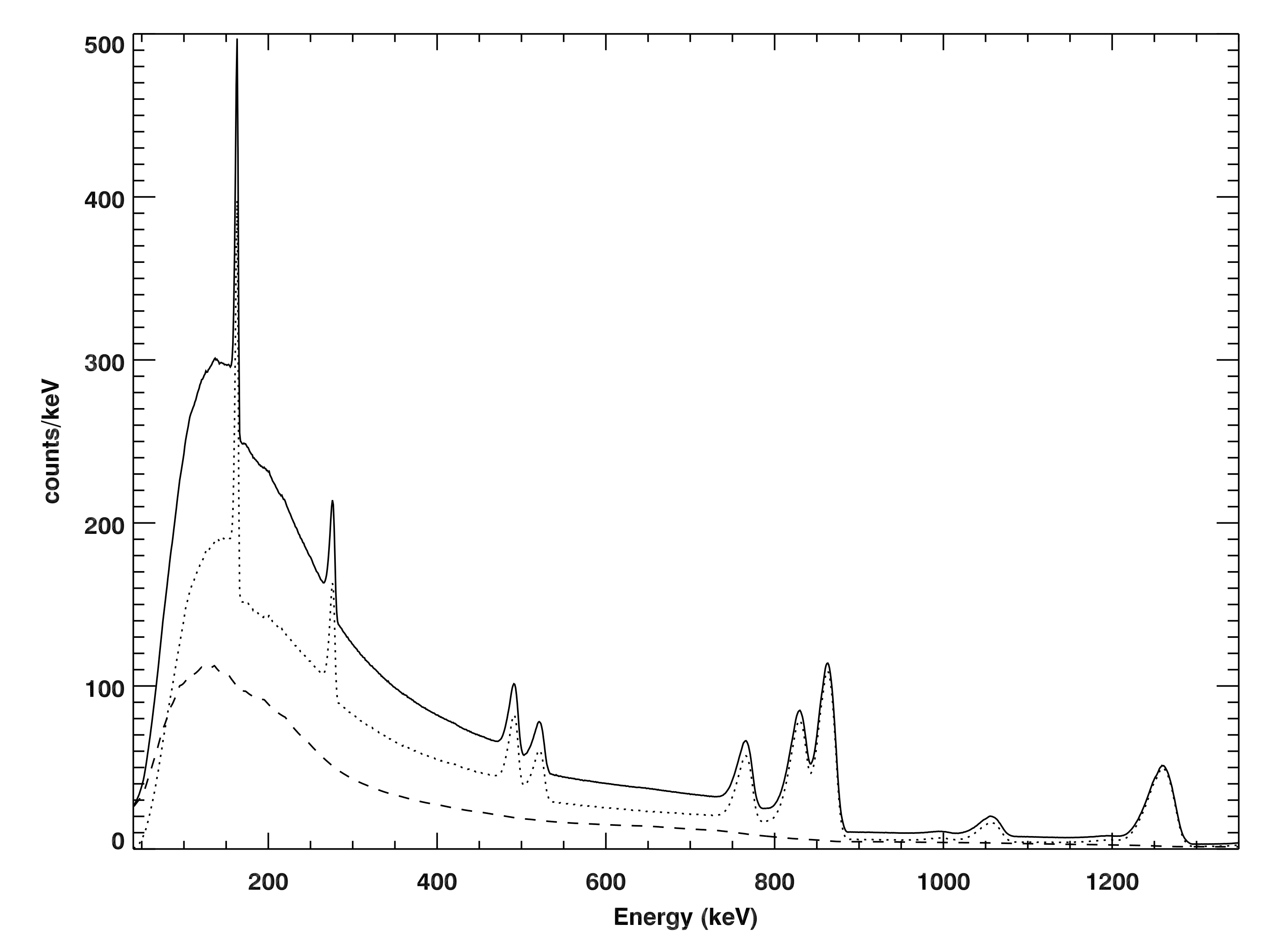}
      \caption{SPI spectral response to the DDT1p4 model during revolutions 1380-86. The dotted and dashed lines represent the contribution of the diagonal and off-diagonal terms of the convolved spectrum respectively. The solid line is the sum of both components. The diagonal component scales with the original model while the non-diagonal component is produced by the energy redistribution of high energy photons and contributes to the continuum with an amount comparable to that of the diagonal component in the 100-200 keV range.
              }
         \label{fig2}
   \end{figure}

In the case of SPI, the convolved spectrum is calculated for a given theoretical model taking into account the SPI IRF (Imaging Response Files) and RMF (Redistribution Matrix Functions)\footnote{ 
For more details on the method, see Compact Source Analysis document: \url{http://www.isdc.unige.ch/integral/download/osa/doc/10.1/spi_compact_source_analysis.pdf}},
 where the RMF were calculated by Monte Carlo simulations \citep{stur03}.
This convolution method has been successfully tested using the data obtained from the Crab Nebula observations during revolution 1387 and the results obtained by \citet{jour09}. The model convolutions were performed with $IRF$s and $RMF$s version 7.0 and the theoretical models used in this work are described in appendix~\ref{stm} and table~\ref{tm}. As an example, Fig.~\ref{fig2} presents the DDT1p4 spectrum convolved with the spectral responses of SPI during revolutions 1380-1386 as well as the influence of the off-diagonal terms.

   \begin{figure}
   \centering
 \includegraphics[width=\hsize]{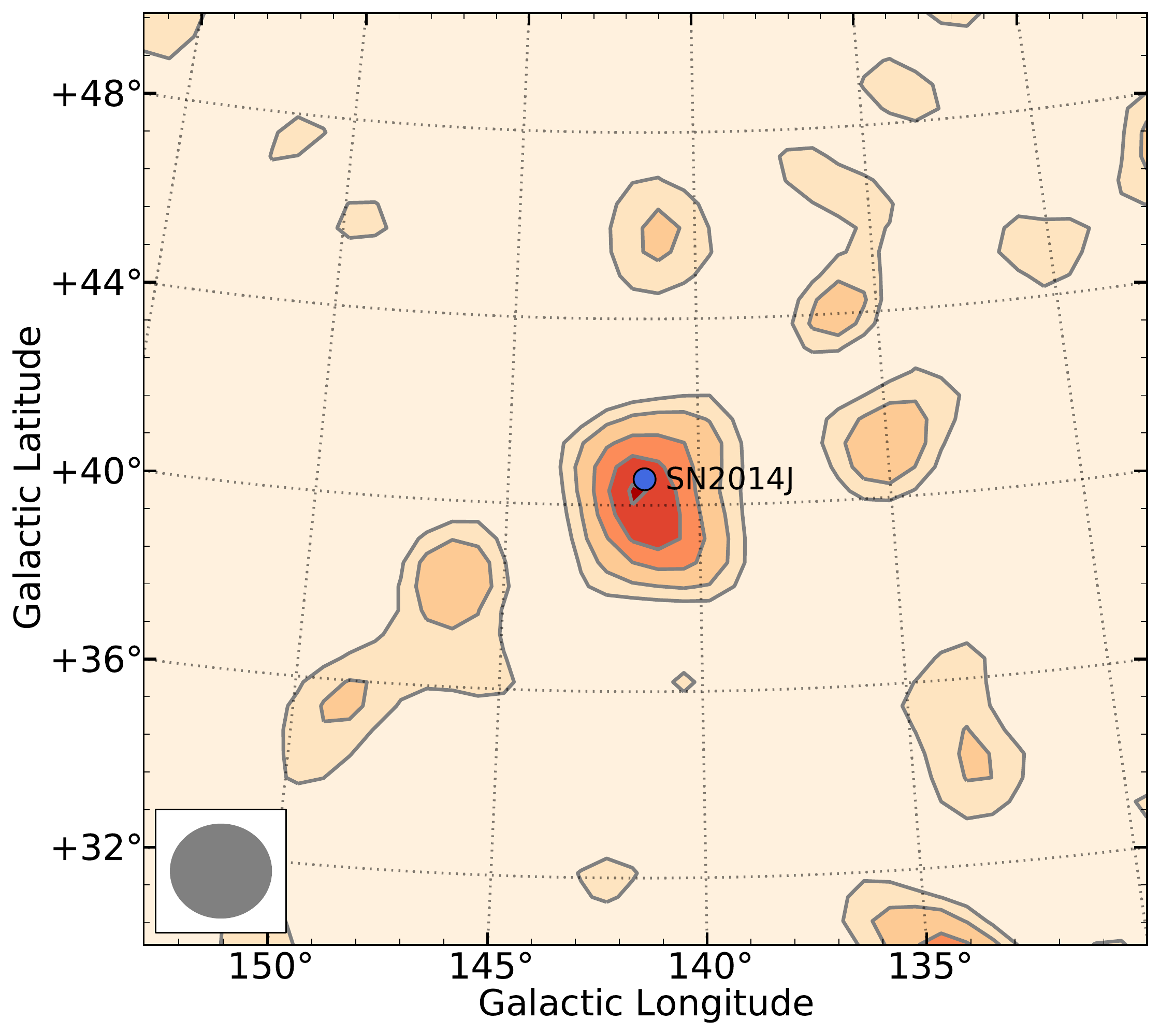}
      \caption{
Gamma-ray signature of SN2014J in the SPI data. The panel displays the statistics map for single detector events (SE) obtained by SPI during the entire early period of observation (days $\sim ~ 16-35$ days after the explosion) over the energy band 145-165 keV. The maximum likelihood ratio of the contour lines is 0, 2.5, 5, 10, 15 and 20. This statistics map has been obtained by fitting a point source on top of the background and the background alone for each position using pixels of 0.5 degrees. The excess in the SN2014J position is $5 \sigma$.
              }
         \label{fig3}
   \end{figure}

 The analysis of the data obtained by SPI during this first observation period has revealed an emission excess in the 70-190 and 650-1300 keV bands at the position of SN2014J that was not present in the observations performed by INTEGRAL before the explosion.  Figure \ref{fig3} displays this emission excess in the energy band of 145-165 keV, where the 158 keV \element[ ][56]{Ni} gamma-ray line is expected to lay. The significance of this excess, $5 \sigma$,  is computed subtracting the log of the maximum likelihood values obtained by fitting both the background alone and the background plus source. The figure also shows that the maximum of the emission coincides with the position of SN2014J, $l=140.5^o$, $b=42.5^o$, and it is clearly isolated from the neighbouring sources as seen by SPI. This localization represents an improvement with respect to the offset of $\sim 2^{\rm o}$ present in the previous values reported by \citet{dieh14}.

\begin{figure}
   \centering
 \includegraphics[width=\hsize]{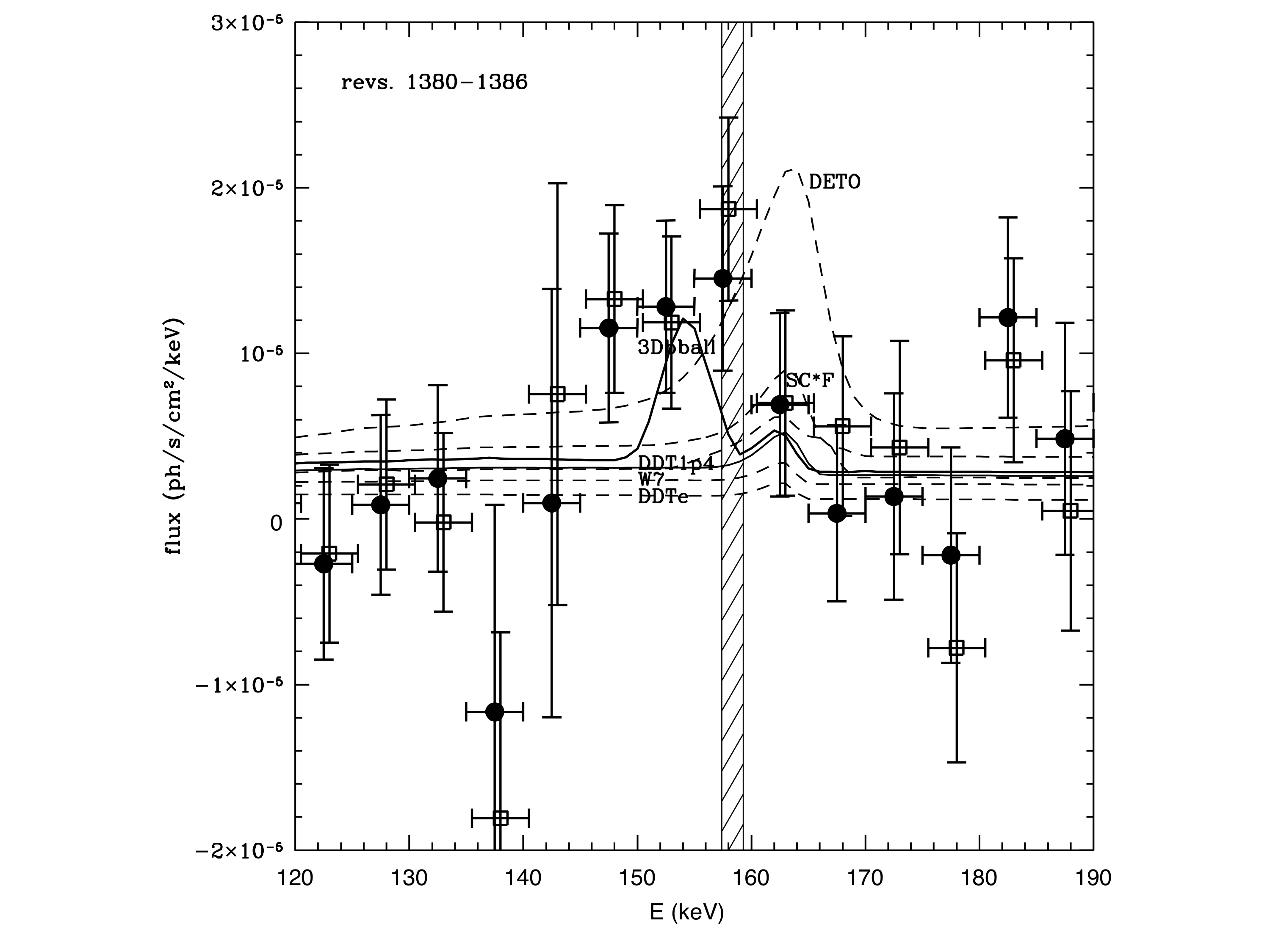}
      \caption{ Spectrum of SN2014J obtained by SPI during revolutions 1380-1386 in bins of 5 keV in the 120-190 keV band. Filled circles were obtained with the procedures described in \citet{iser13} and empty squares as in \citet{chur14a}. The last ones have been shifted 0.5 keV for a sake of clarity. The lines represent the signal that is expected from a subset of theoretical models listed in Table~\ref{tm} (SC*F means SC1F and SC3F) after convolving with the SPI response.  The shaded region is centred at the nominal energy of the \element[ ][56]{Ni} line, 158.4 keV, and its width is equal to the energy resolution of SPI at this energy. 
              }
         \label{fig4}
   \end{figure}

In the low energy region, it has been found in the SPI data a broad and completely unexpected redshifted feature associated to the 158 keV \element[ ][56]{Ni} gamma-ray line\footnote{ See however \citet{dieh14} for a different analysis.}. Figure~\ref{fig4} displays the spectrum obtained during orbits 1380-1386 (16.5-35.2 days after the explosion) by SPI in the 120-190 keV band using two independent procedures and the spectrum predicted by different theoretical models (see Appendix \ref{stm}) after being convolved with the SPI response.  Notice that all the classical, spherically symmetric models predict a blueshifted line at this epoch and that the continuum, which depends on the adopted model, is in the range of $1.5 \times 10^{-6}$ and $5 \times 10^{-6}$ ph cm$^{-2}$s$^{-1}$keV$^{-1}$.  Figure \ref{fig4} also shows the concordance between the two independent analysis of data that have been performed in the region where the 158 keV \element[ ][56]{Ni} line should be placed. 

 \begin{figure}
   \centering
 \includegraphics[width=\hsize]{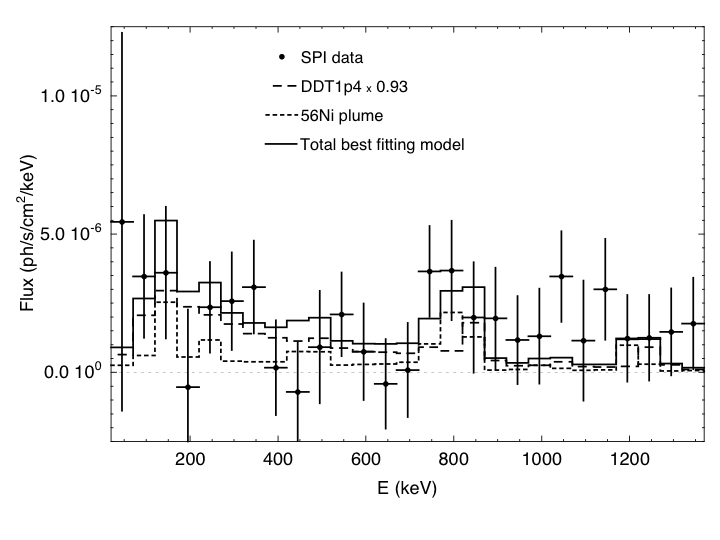}      
  \caption{ Gamma-ray spectrum during revolutions 1380-1386 (16.5-35.2 days after the explosion). Bins are 50 keV wide. The continuous line represents the best fit obtained scaling the model DDT1p4 by a factor 0.93 (0.605 M$_\odot$ of \element[ ][56]{Ni}) - long dashed line- and adding a \element[ ][56]{Ni} plume of 0.077 M$_\odot$-short dashed line.  
              }
         \label{fig5}
   \end{figure}

\begin{figure}
   \centering
 \includegraphics[width=\hsize]{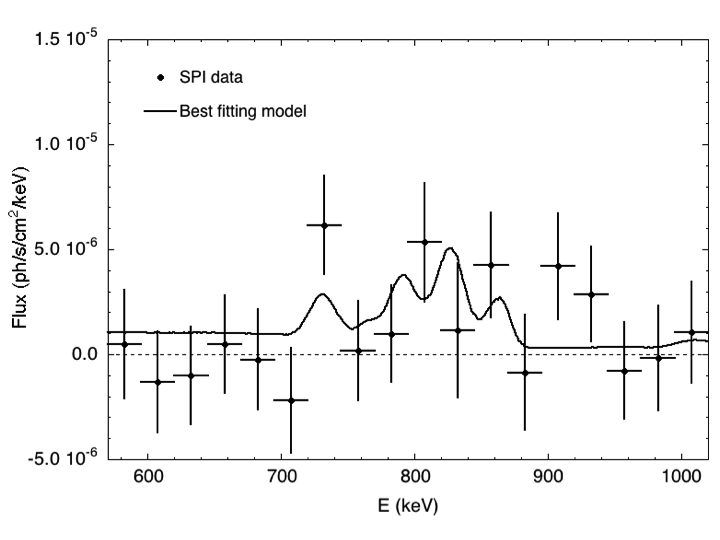}      
  \caption{ Gamma-ray spectrum during revolutions 1380-1386 (16.5-35.2 days after the explosion) but with bins of 25 keV size. As in Figure~\ref{fig5}, the continuous line represents the best fitting model convolved with the SPI response.  
              }
         \label{fig6}
   \end{figure}

Figure~\ref{fig5} presents the spectrum from 20 keV to 1370 keV with a binning of 50 keV, where a flux excess in the 720-870 keV energy band can also be seen. The significance of this excess is $\sim$ 2.8 $\sigma$ and can be attributed to the contribution of the \element[ ][56]{Ni} and \element[ ][56]{Co} decays. Unfortunately, the blending of the 812 keV \element[ ][56]{Ni} and 847 keV\element[ ][56]{Co} lines, caused by the Doppler broadening \citep{gome98} together with the relative weakness of the fluxes, prevents any spectroscopic analysis of these individual gamma-ray lines. It is also interesting to notice the presence of a feature  with a 2.6 $\sigma$ significance at  $\sim 730$ keV, the position that would correspond to the 750 keV \element[ ][56]{Ni} line  redshifted by the same amount as the 158 keV line (see Figure~\ref{fig6}). The gaussian fit of this feature gives  a flux of $(1.5\pm 0.7) \times 10^{-4}$ ph s$^{-1}$cm$^{-2}$ (2.1 $\sigma$), a centroid placed at $733.4 \pm 3.8$ keV and a FWHM of $16.9 \pm 9.0$ keV. 

If it is assumed that the redshifted feature associated to the 158 keV line is due to \element[ ][56]{Ni}, the other gamma-ray lines emitted by this isotope should also be redshifted and their widths and fluxes should be in agreement with those of the 158 keV line taking into account their branching ratio (see below). Therefore, the inclusion of the measured bins of the high energy lines in the spectral analysis provides an additional constraint to the analysis of the \element[ ][56]{Ni} emission. Consequently, the flux, the width and the redshift of the 158 keV line were fitted to the data by linking these three parameters to the respective fluxes, broadening and redshifts of the 750 keV and 812 keV lines with their corresponding branching ratio (0.50 and 0.86, respectively). Under these conditions, the best fit with a gaussian that links this feature with the red-shifted 750 and 812 keV \element[ ][56]{Ni} lines gives a flux of $(1.6\pm 0.4) \times 10^{-4}$ cm$^{-2}$s$^{-1}$, centred at $155.2 ^{+1.3}_{-1.1}$ keV with a FHWM $5.2^{+3.4}_{-2.2}$ keV.  These values were obtained from the analysis of the 2 keV bin spectrum (615 bins) between 120 keV to 1350 keV. Taking into account there are three free parameters, energy shift, broadening and flux in the 158 keV line, this gives $\chi^2 = 564.76$ and a reduced $\chi^2= 0.923$ for a dof 612. The null hypothesis yields to a $\chi^2= 592.251$ - i.e. the $\Delta \chi^2 \sim 27.5$. 

These results contrast with those found by \citet{dieh14},  who observed two very narrow lines placed at the nominal values of the 158 and 812 keV  \element[ ][56]{Ni} features, very near to the aforementioned instrumental lines. 
 Fortunately, the 158 keV feature found in this work is broad and is shifted to $\sim 155$ keV, where there are no such background lines.

\begin{figure}
   \centering
\includegraphics[width=\hsize,angle=90]{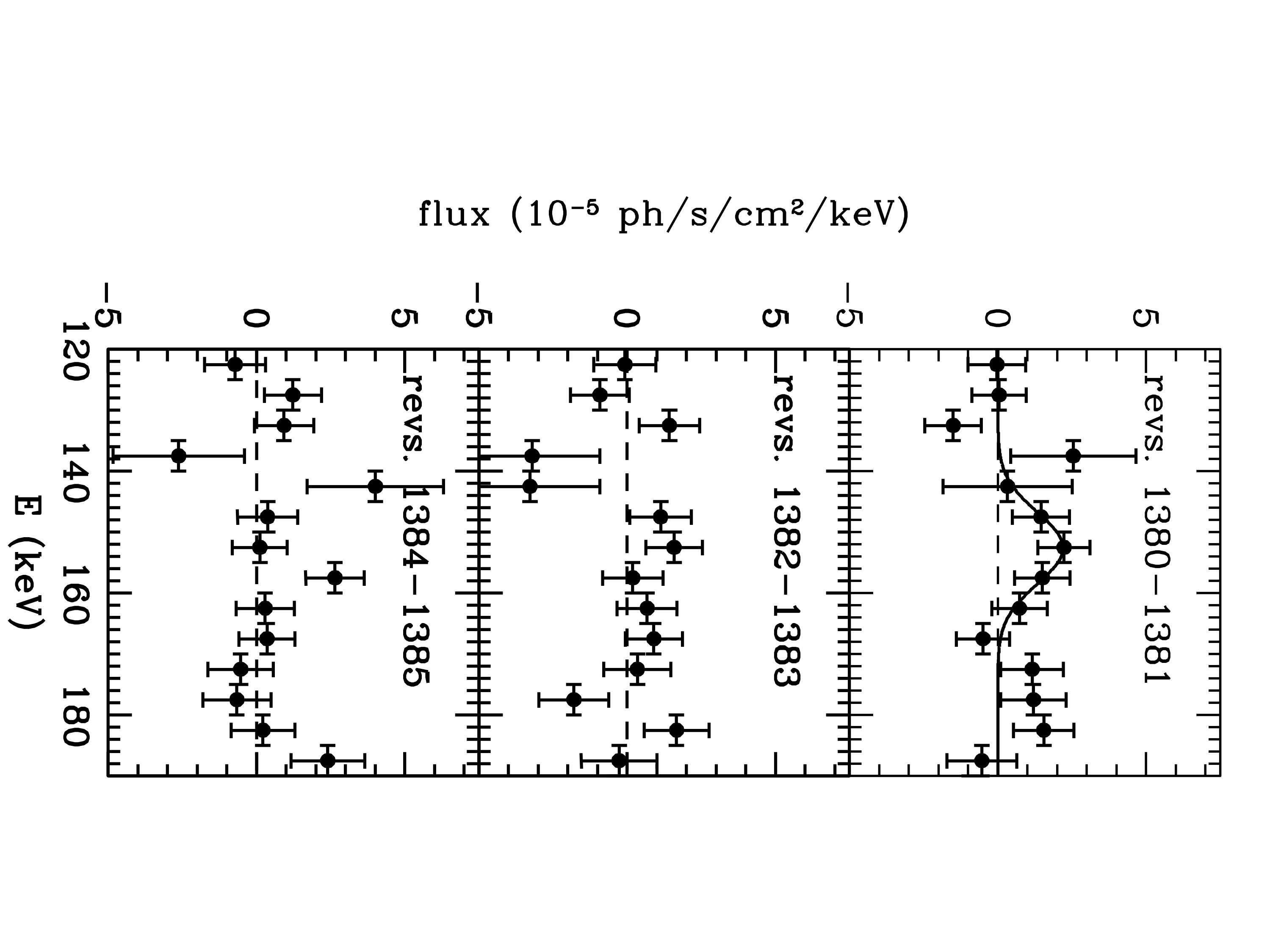}
      \caption{Spectral evolution of SN2014J obtained by SPI in the 120-190 keV band during revolutions 1380-81, 1382-83 and 1384-85, or 16.5-22.2, 22.6-28.2, and 28.6-35.2 days after the explosion (top to down) respectively, extracted in bins of 5 keV. 
              }
\label{fig7}
\end{figure}

The evolution of the spectrum during this early phase of observation could also provide some hints on the nature of the explosion. With such a purpose, data in the 120-190 keV band were grouped into bins corresponding to revolutions 1380-81, 1382-83 and 1384-85. These time intervals were chosen as a compromise between an optimal signal to noise ratio and the possibility to solve in time the light curve. 
Figure~\ref{fig7} displays the gaussian fits obtained in this way. The flux measured during revolutions 1380-1381 is $(2.23 \pm 0.80) \times 10^{-4}$  ph cm$^{-2}$ s$^{-1}$, centered at $152.6 \pm 2.8$ keV and with a significance of 2.8 $\sigma$. In the other two bins the signal to noise ratio is too poor to perform any definite comparison about the evolution of the lines and only upper limits ( 2 $\sigma$ level) can be provided: $ < 1.72 \times 10^{-4}$, and $< 1.53 \times 10^{-4}$  ph cm$^{-2}$ s$^{-1}$. In any case, these values are an overestimation of the flux since the intrinsic continuum of SN2014J, detected during the late observations \citep{chur15} and the complete spectral response of SPI were not taken into account. 

  If the intrinsic continuum of SN2014J is taken into account and the complete response of SPI is adopted, the fluxes in the gaussian fits of the spectra presented in Fig.~\ref{fig7} become $(1.59 \pm 0.57) \times 10^{-4}$, $ < 1.42 \times 10^{-4}$ and $<1.56 \times 10^{-4}$ photons cm$^{-2}$$s^{-1}$ after removing the continuum underneath the line produced by the off-diagonal terms of the DDT1p4 model. In this case, the most significant gamma-ray line signature from the \element[ ][56]{Ni} occurs during revolutions 1380-1381 with a centroid at $154.5 \pm 0.64$ keV and a width of $3.7 \pm 1.5$ keV.

\subsection{IBIS/ISGRI data}

\begin{figure}
   \centering
\includegraphics[width=\hsize]{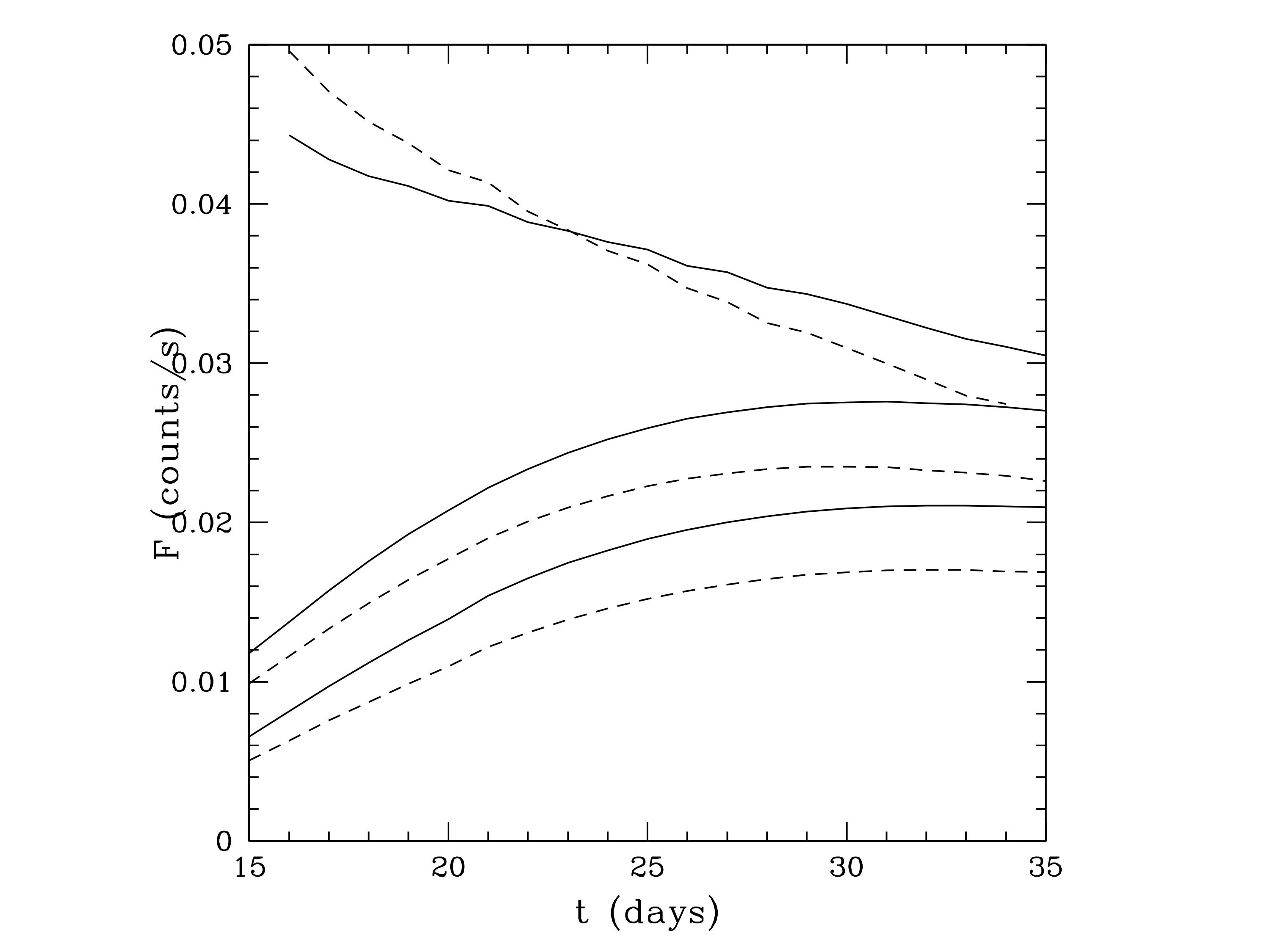} 
      \caption{IBIS/ISGRI light curve in response to different spectral models. Dashed lines represent the temporal evolution of the  144.5 - 168 keV band of the 3Dbball, DDT1p4 and W7 models (from top to bottom)  multiplied by the effective area in this band (370 cm$^2$).  Continuous lines represent their convolution with the ISGRI response.
              }
         \label{fig8}
   \end{figure}

As in the case of SPI, the data obtained by IBIS/ISGRI have been analyzed independently with the method described in \citet{iser13}, which takes into account the response of the instrument, but using the OSA-10 instead of OSA-9 since it noticeably improves the reconstruction of the photon energy, and with the method described in \citet{chur14a}, where the flux is obtained by normalizing to the values of the Crab in the same energy band.  Usually, this normalization procedure is sufficient if the energy band being analyzed is broad enough, but this is only strictly valid if both spectra, Crab and supernova, were similar, which is not the case. For this reason, the procedure adopted here is to compare the observations with the theoretical models convolved with the ISGRI spectral response.  The energy resolution at 155 keV was FWHM $\sim$ 14 keV, but due to the detector degradation in orbit, the resolution is now closer to 20\% \citep{caba12}.  In order to show how the evolution of the count rate depends on the spectrum of the adopted models and how they evolve with time,  Figure~\ref{fig8} displays the count rate that is obtained  from three of the models used here (continuous lines after convolving with the response of the instrument, and the count rate obtained just multiplying the theoretical fluxes by the effective area in the energy band under consideration. 

\begin{figure}
   \centering
 \includegraphics[width=\hsize]{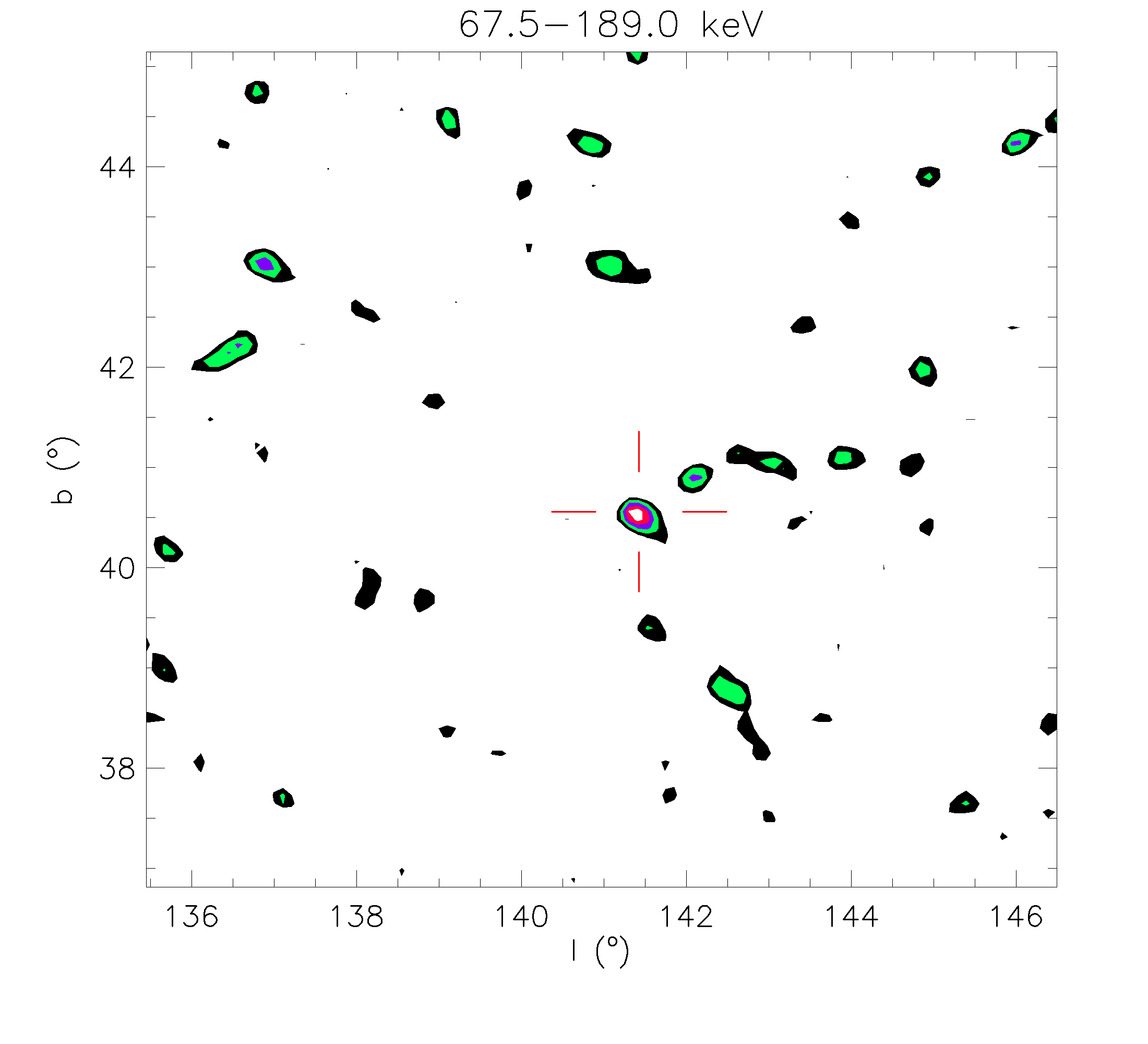}
 \includegraphics[width=\hsize]{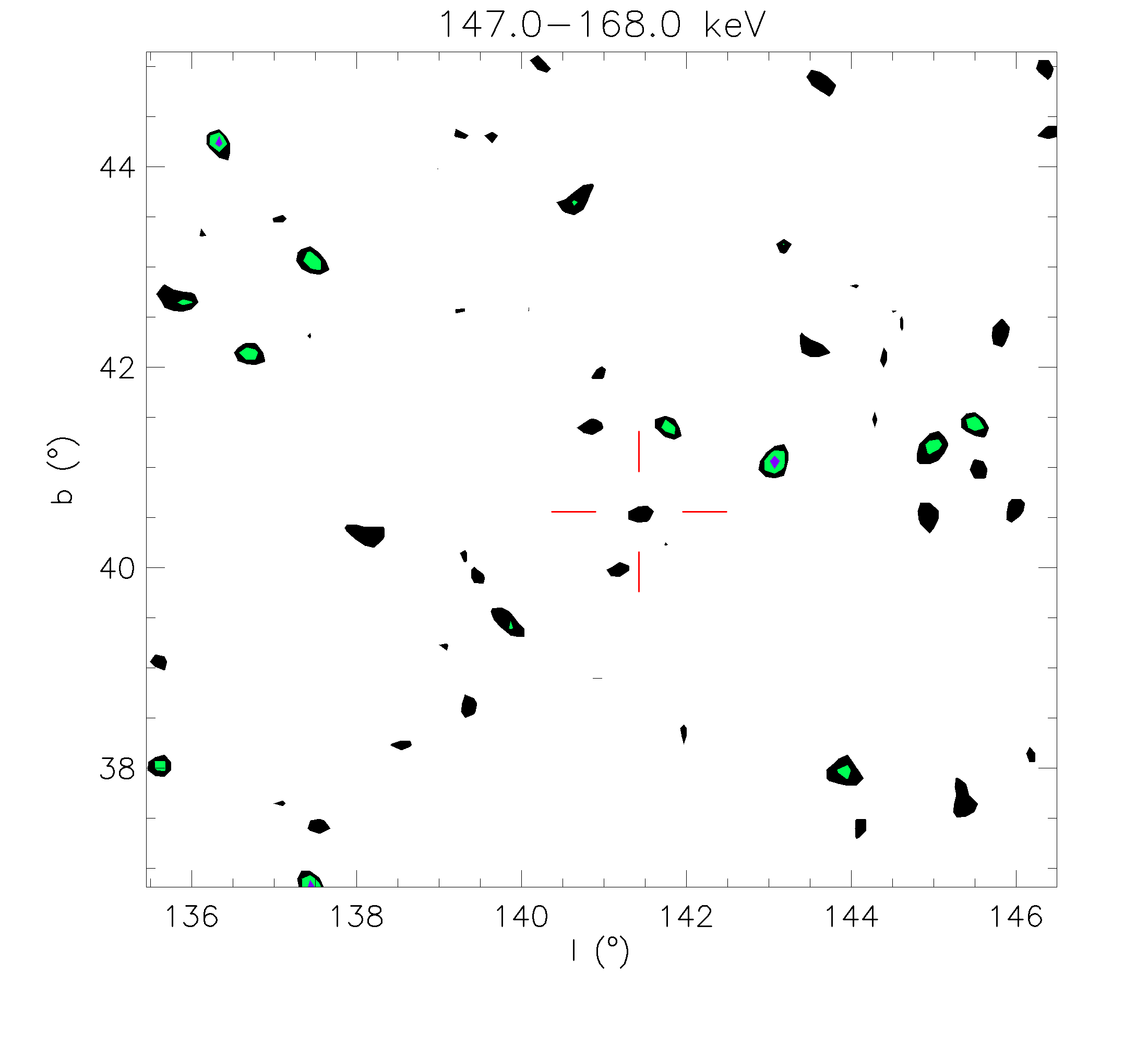}
      \caption{
Gamma-ray signature in the IBIS/ISGRI data. The figure displays the IBIS/ISGRI significance contour map in the 67.5 – 189 keV band (upper panel) and in the 147 - 168 keV band (lower panel) for the entire early period (days  ~ 16 - 35 after the explosion).  The contours start at the 2 $\sigma$ level and are separated by 0.5 $\sigma$. The average flux in the upper panel at the position of SN2014J represents a $5.4 \sigma$ excess after normalization on the standard deviation observed in this map. 
              }
         \label{fig9}
   \end{figure}

Taking into account the ISGRI spectral response, the signal expected from most models is maximum over the 68-190 keV range, and the observations performed by IBIS/ISGRI during the same period of time as SPI reveal an emission excess at the position of SN2014J in the energy band 67.5-189 keV. Figure~\ref{fig9}, upper panel, clearly shows that this emission excess cannot be confused with the neighbouring sources.  However, if the analysis is restricted to the 144.4 – 168 keV band, the significance of the signal decreases to about 2 sigma, as expected from the ISGRI spectral response (see Fig. \ref{fig9}, lower panel). In the 25 - 70 keV band nothing is visible at the position of SN2014J.

\begin{figure}
\centering
\includegraphics[width=\hsize]{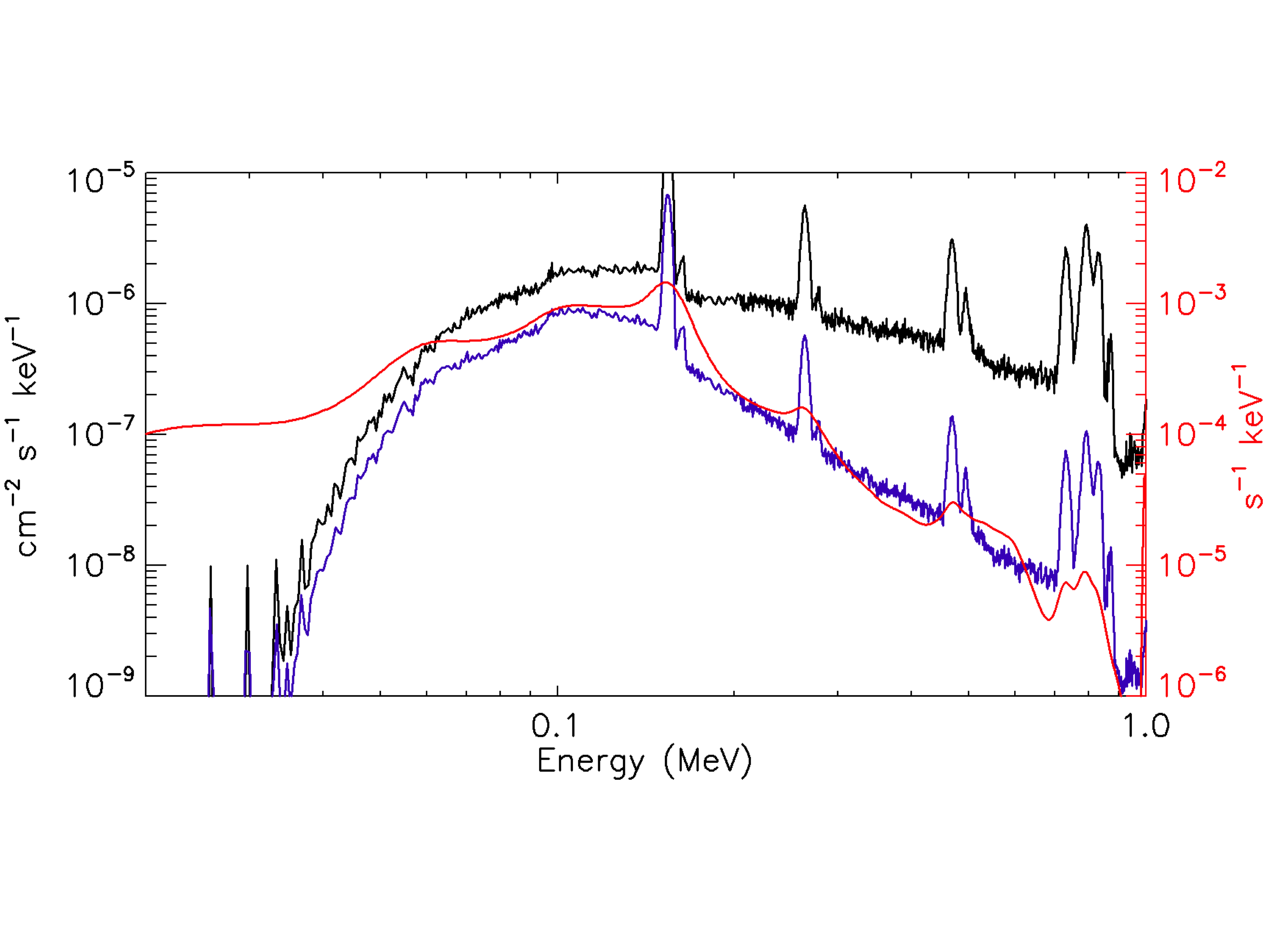}
\caption{
Response of IBIS/ISGRI to an incoming gamma signal that has a characteristic supernova spectrum like the one provided by the 3Dbball model (black line). The blue line is obtained multiplying this spectrum by the ARF (Auxiliary Response Files) that represent the effective area, and finally the red line represents the values obtained by convolving this last result with the RMF (Redistribution Matrix Function).
}
\label{fig10}
\end{figure}

 Figure~\ref{fig10} displays the response  of ISGRI to an incoming gamma-ray flux that has the same spectrum as one of the models used in this work, the 3Dbball model. As it can be seen, the photons belonging to the 158 keV \element[ ][56]{Ni} lines are redistributed over a spectral band that is larger than expected. As a consequence, the flux of the line weakens when a narrow spectral window is taken.

   \begin{figure}
   \centering
 \includegraphics[width=\hsize]{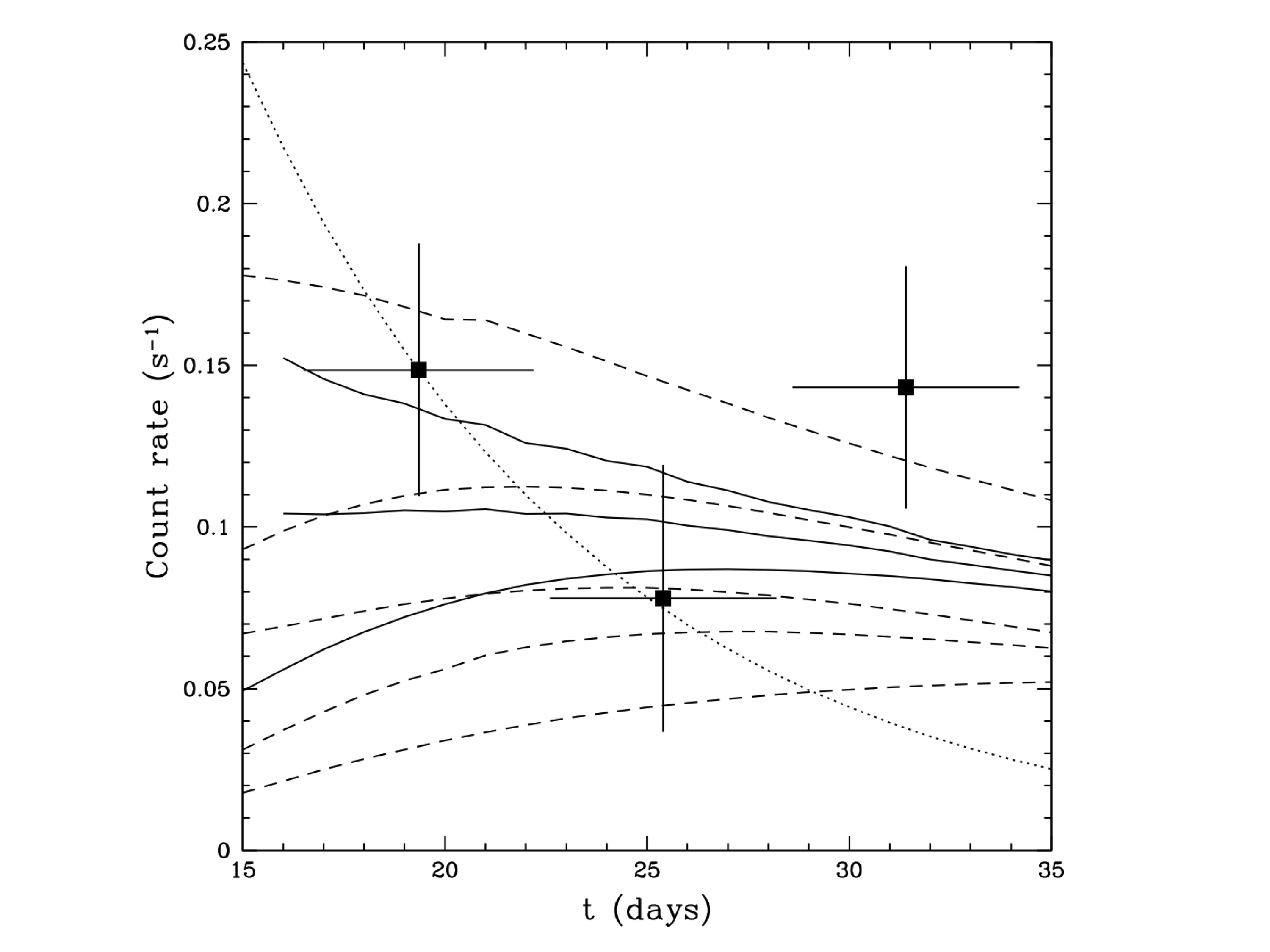}
      \caption{Evolution of the {bf 67.5 - 189} keV band during revolutions 1380-81, 1382-83 and 1384-86 as obtained by IBIS/ISGRI using the method described in \citet{iser13}. The dotted line represents the flux that would be provided by the free disintegration of \element[ ][56]{Ni}. Dashed lines represent the light curves obtained by convolving models DETO, SC3F, SC1F, W7 and DDTe (from top to down) with the instrument response. Solid lines represent, from top to down, the 3Dbball with plumes of 0.08 and 0.04 M$_\odot$ of \element[ ][56]{Ni}, and the DDT1p4 models after convolution. The properties of these models are displayed in Appendix~\ref{tm}.  
              }
         \label{fig11}
   \end{figure}

Given the strong redistribution of photons, only the broad band of 67.5 - 189 keV will be considered. Table~\ref{tisgri} and Figure~\ref{fig11} display the temporal evolution of the count rate measured by IBIS/ISGRI in this band, which is dominated by the 158 keV \element[ ][56]{Ni} emission. As in the case of SPI, bins are roughly six days wide and correspond to revolutions 1380-81, 1382-83 and 1384-85. The behavior, similar to that obtained by SPI but with a better significance, suggests a decline in the count rate that is compatible with the non-absorbed emission of \element[ ][56]{Ni}, followed by an upturn at the end of the observation period that could be the consequence of the exposure of new radioactive layers. Unfortunately,  the poor S/N ratio of the central bin (1.9 $\sigma$) prevents any solid conclusion about this point, and an approximately constant or gently decaying behavior cannot be excluded. 

\begin{table}
\caption{Temporal evolution of the 158 keV \element[ ][56]{Ni} according to IBIS/ISGRI.}
\label{tisgri}
\centering
\begin{tabular}{c c c c}
\hline\hline
\\
Revolutions & Days & counts/s & S/N \\
\hline
1380-1381 & 22.20-16.50 & 0.149 $\pm$ 0.039 & 3.8 \\
1382-1383 & 28.20-22.60 & 0.078 $\pm$ 0.041 & 1.9 \\
1384-1385 & 34.20-28.60 & 0.143 $\pm$ 0.037 & 3.8 \\
\hline
\end{tabular}
\end{table}

\section{Results and discussion}
\label{secrd}

There are several spherically symmetric SNIa models (see Appendix~\ref{stm}) with the bulk of radioactive elements buried in the central layers of the expanding debris that are able to reproduce, with the appropriate parameters, the \element[ ][56]{Co} features observed at late times, 55-100 days after the explosion, in SN2014J \citep{chur14a,chur15}. These models, after being convolved with the SPI response, as in the case of DDT1p4 presented in Figure~\ref{fig5}, can be compared with the observed spectrum taken during revolutions 1380 to 1386, in the range 120-1350 keV. The degrees of freedom are 246, and Table~\ref{tmchi} presents the resulting $\chi^2$ values. 

The DDT1p4 model explains the optical light curve (see section 2) and the gamma-ray emission at late epoch (e.g. Churazov et al. 2014). If, in order to make a crude comparison and using the same criteria as in \citet{chur15}, we adopt this model as a reference we see that DETO and DDTe differ by $\sim 6 \sigma$ and $\sim 3 \sigma$ level respectively, while the remaining ones are nearly as good as the DDT1p4 model ($< 1.5 \sigma$). However, despite the reasonable agreement with the observed values of these remaining models, they are neither able to reproduce the intensity and the redshifted position of the 158 keV \element[ ][56]{Ni} line observed by SPI (Fig. \ref{fig4}) nor the excess of emission in the bin corresponding to the 1380-81 orbits found in the IBIS/ISGRI data (Fig.~\ref{fig11}) and, in a less compelling form, by SPI (Fig.~\ref{fig7}). Only DETO seems to fulfill such last requirements but it synthesizes a total amount of \element[ ][56]{Ni} that is too large to account for the late emission of \element[ ][56]{Co}. It also predicts the presence of important amounts of this isotope in the outer layers that is in contradiction with the optical observations during the maximum of the light curve.
Therefore, it seems  natural to propose models with small amounts of radioactive material in the outer layers of the supernova debris \citep{burr90,gome98} that are undetectable at the other wavelengths.

\begin{table}
\caption{Comparison between models and the spectra measured by SPI (d.o.f.= 246). H$_0$ represents the nulle hypothesis }
\label{tmchi}
\centering
\begin{tabular}{c c}
\hline\hline
Model       &     $\chi^2 $	\\
\hline
H$_0$             &        250.1         \\
DETO        &	269.2         \\
W7            &	229.5          \\
DDTe	        &  	237.0         \\
SC1F	        & 	226.4         \\
SC3F	        &        227.9          \\
\hline
DDT1p4    & 	227.2          \\
3Dbball     & 	220.8           \\
\hline
\end{tabular}
\end{table}

The average 3.2 keV redshift, $\sim ~ 2\%$ of the nominal 158 keV energy, indicates that the material is receding from the observer with a mean velocity $  v \approx 6,000$ km/s and is placed in the far hemisphere, while the measured average width, 4.9 keV, implies a maximum deviation of the component of the velocity along the line of sight of $\Delta v \approx 10,000$ km/s. Nothing can be said about the velocity in the plane normal to the line of sight except that, in order to not be caught by the outer layers of the supernova, it must have a velocity of the order of $ \sim 30,000$ km/s. This possibility could be supported by the rapid rise of the optical light curve at early times \citep{zhen14,goob14} and by the microvariability found 15-18 days after the maximum in the B-light curve \citep{bona15} in SN2014J, by the chemical inhomogeneities found in Kepler \citep{reyn07} and Tycho \citep{vanc95,warr05} remnants, and by the properties of the high velocity features detected in the early optical spectra, $\sim 10$ days after the explosion, of many SNIa \citep{tana06}.

  \begin{figure}
   \centering
 \includegraphics[width=\hsize]{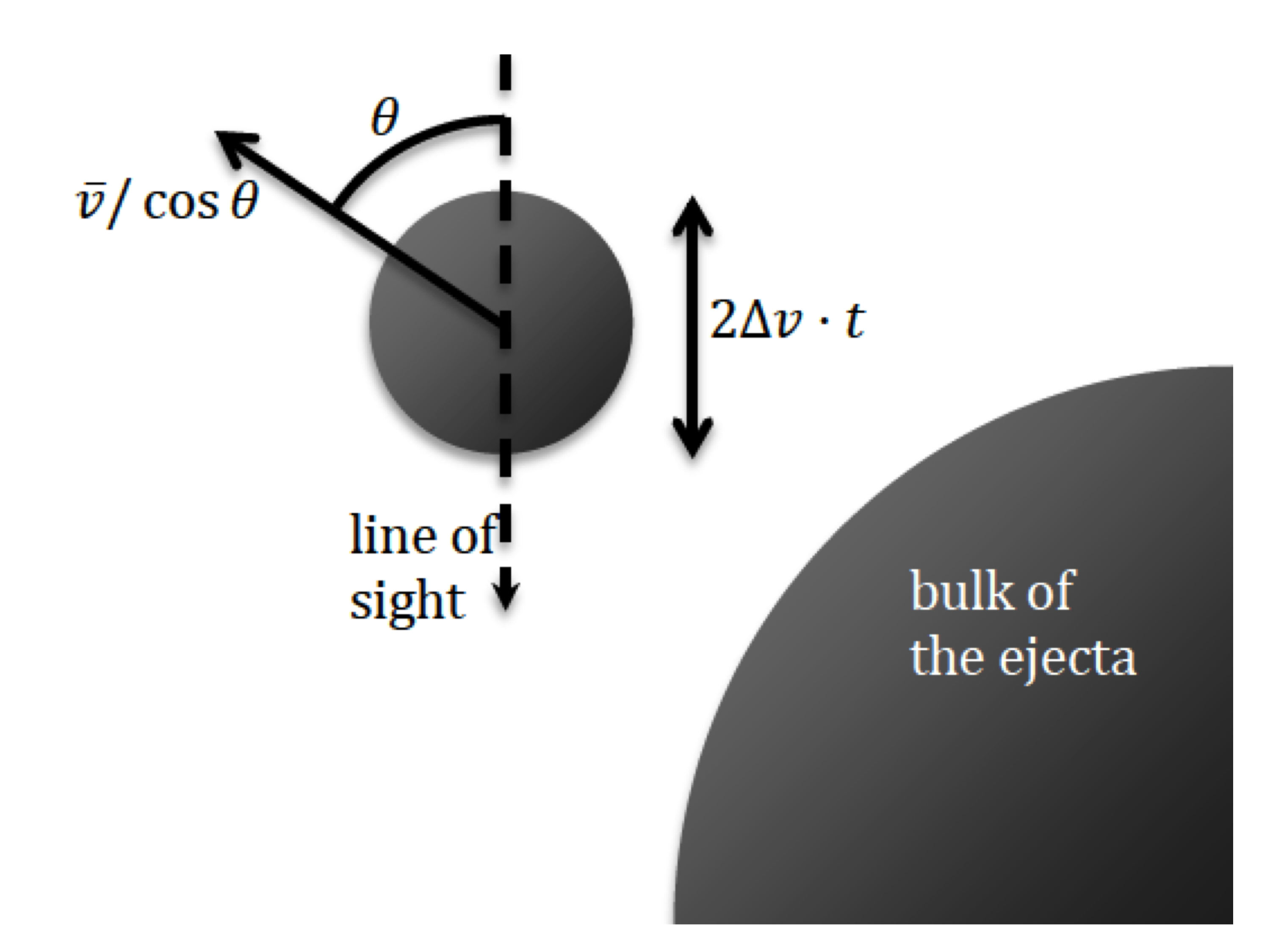}   

\includegraphics[width=\hsize]{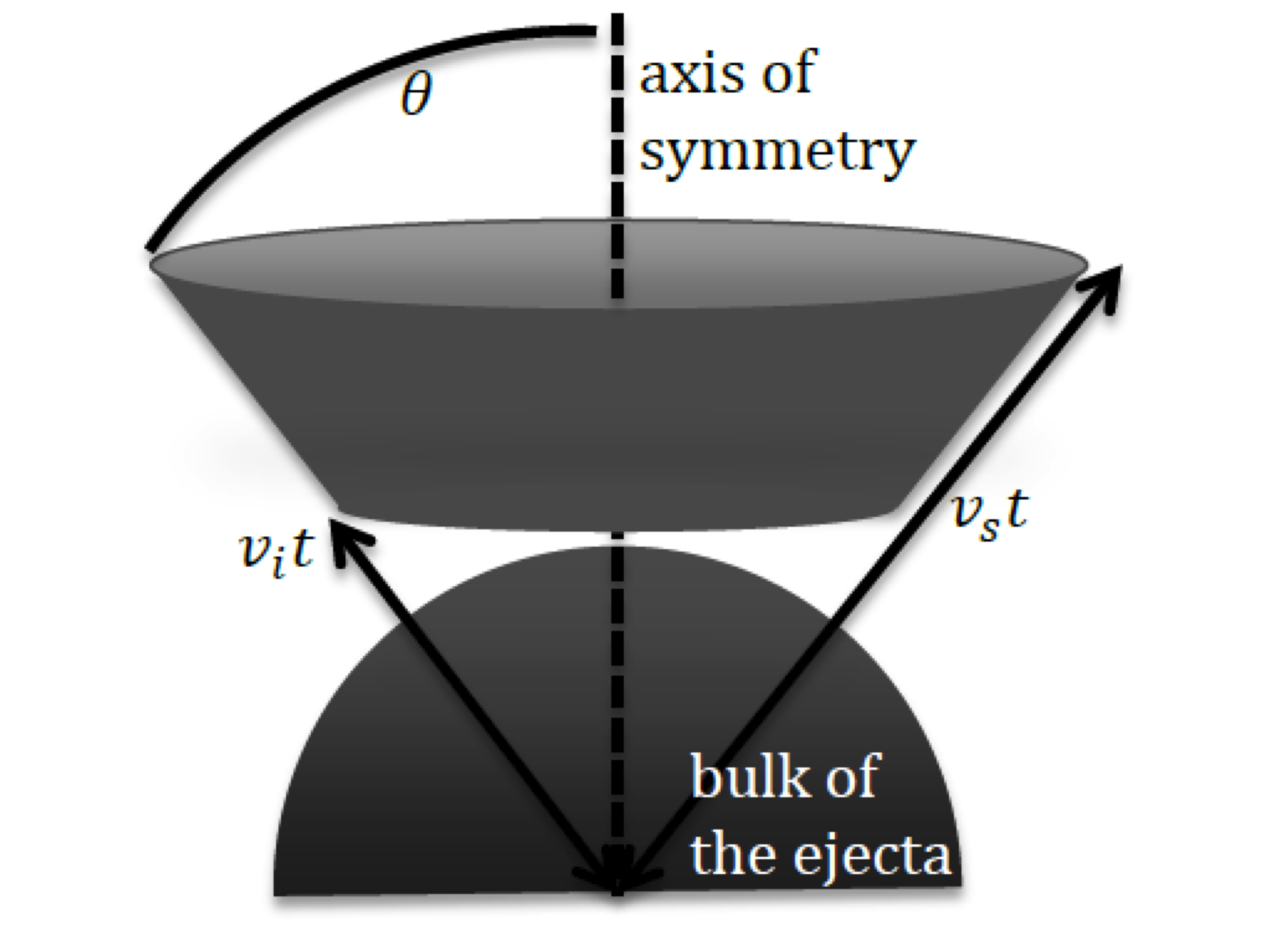}    
  \caption{  Possible geometries of the outer radioactive layers. The upper panel represents a blob with a mass $M \sim  0.05$ M$_\odot$ that detached form the main body during the explosion and moves with a velocity compatible with the observed redshift. The lower panel represents a plume with the shape of a truncated conical ring that has a semiaperture angle $\theta$ and angular thickness $\Delta \theta$.
              }
         \label{fig12}
   \end{figure}

 This radioactive material should be almost transparent to gamma-rays at the time of the \emph{INTEGRAL} observation, near the maximum of the optical light curve, otherwise it would have been detected in the optical. This condition is also necessary to account for the dip suggested by IBIS and SPI data about 25 days after the explosion (figures~\ref{fig11} and \ref{fig7}). Furthermore, the non detection of blue shifted Ni-Co features in the optical and in the infrared at this epoch indicates that it was not placed between the observer and the supernova.  An additional argument in favour of the transparency hypothesis is that an opaque plume would demand a larger \element[ ][56]{Ni} mass to obtain a similar flux and this would introduce a redshifted component in the late \element[ ][56]{Co} emission that is not observed \citep{chur14a,chur15}.
 
The first obvious geometry choice to be considered is a spherical blob that broke away from the bulk of the supernova ejecta. Such a configuration, however, does not guarantee the transparency of the blob at the moment of the observation. For instance,  Figure~\ref{fig12}, upper panel, displays a blob with a mass $M \sim 0.05$ M$_\odot$, expelled by the main body of SN2014J with a velocity of $\sim 30,000$ km/s and increasing its radius with a velocity of the order of $\Delta v \sim 10,000$ km/s, the internal velocity dispersion. At day 18 after the explosion, the radius of the ball should be $\Delta v \times t =1.5 \times10^{15}$ cm and the optical depth $\tau = \kappa \Sigma \approx  3$, where $\Sigma \sim  M/\pi R^2 \approx 14$ g cm$^{-2}$. Assuming constant density, and $\kappa \approx 0.2$ cm$^2$g$^{-1}$ for the 158 keV line, the blob would be opaque to the gamma-ray radiation at the moment of the observation and should be detectable in the optical.

A more favourable geometry is to distribute the radioactive matter in a ring with a truncated conical shape as depicted in Figure~\ref{fig12}, lower panel. We call this model 3Dbball. This conical structure has the same mass as before and, in order to be compatible with the observations of SN2014J, a semi-aperture angle $\theta \approx 78^o$, an angular thickness $\Delta \theta \approx 12^o$ were adopted as an example, although other possibilities do exist. Similarly, the expansion velocities were set to be $v_{\mathrm{i}} \sim 25,000$ km/s and $v_{\mathrm{s}} \sim 35,000$ km/s in order to fulfill the requirements. In this case, the column density would be $\Sigma = M/4\pi (R_{\mathrm{s}} - R_{\mathrm{i}})^2\sin \theta \sim 2.4$ g cm$^{-2}$, where $R_{\mathrm{s,i}} = v_{\mathrm{s,i}} t$, and the optical depth would be 0.48, which would make the material of the plume optically thin to the 158 keV photons. 

Assuming the complete transparency to gamma-rays, the total mass of \element[ ][56]{Ni} can be estimated to be
\begin{equation}
\begin{array}{c c l}
m_0  & = & 4\pi D^2 \left( {\frac{{56}}{{N_A }}} \right)\left[ {\frac{{\left\langle {F_{158} } \right\rangle  \left( {t_i  - t_0 } \right)}}{{Y_{158} }}} \right] \\
         & \times & \left[ {\exp \left( { - {{t_0 } \mathord{\left/{\vphantom {{t_0 } {\tau _{Ni} }}} \right.
 \kern-\nulldelimiterspace} {\tau _{Ni} }}} \right) - \exp \left( {{{ - t_i } \mathord{\left/
 {\vphantom {{ - t_i } {\tau _{Ni} }}} \right.
 \kern-\nulldelimiterspace} {\tau _{Ni} }}} \right)} \right]^{-1}
\end{array}
\end{equation}
\noindent
where $N_{\rm A}$ is the Avogadro’s number, $D$ the distance, $Y_{\rm 158}$ the branching ratio, $t_0$, $t_{\rm i}$ the beginning and the end time of the observation, $\tau_{\rm Ni}$ the characteristic decay time and $\langle F_{\rm 158} \rangle$ the average flux in this time interval. 
The flux measured by SPI during revolutions 1380-1381 indicates that the mass in the plume should be $\sim 0.07$ M$_\odot$ while the mass necessary to account for the flux measured by ISGRI, when the deconvolution method is used, is $\sim 0.08$ M$_\odot$. These values seem to give support to the hypothesis that \element[ ][56]{Ni} is present in the outer layers. There is also a hint, provided by the increase of the intensity of the line flux during revolutions 1384-1385, of the emergence of a new, deeper, plume or of the exposure of the internal radioactive core, but the lack of enough significance of the signal prevents any firm conclusion.

  \begin{figure}
   \centering
 \includegraphics[width=\hsize]{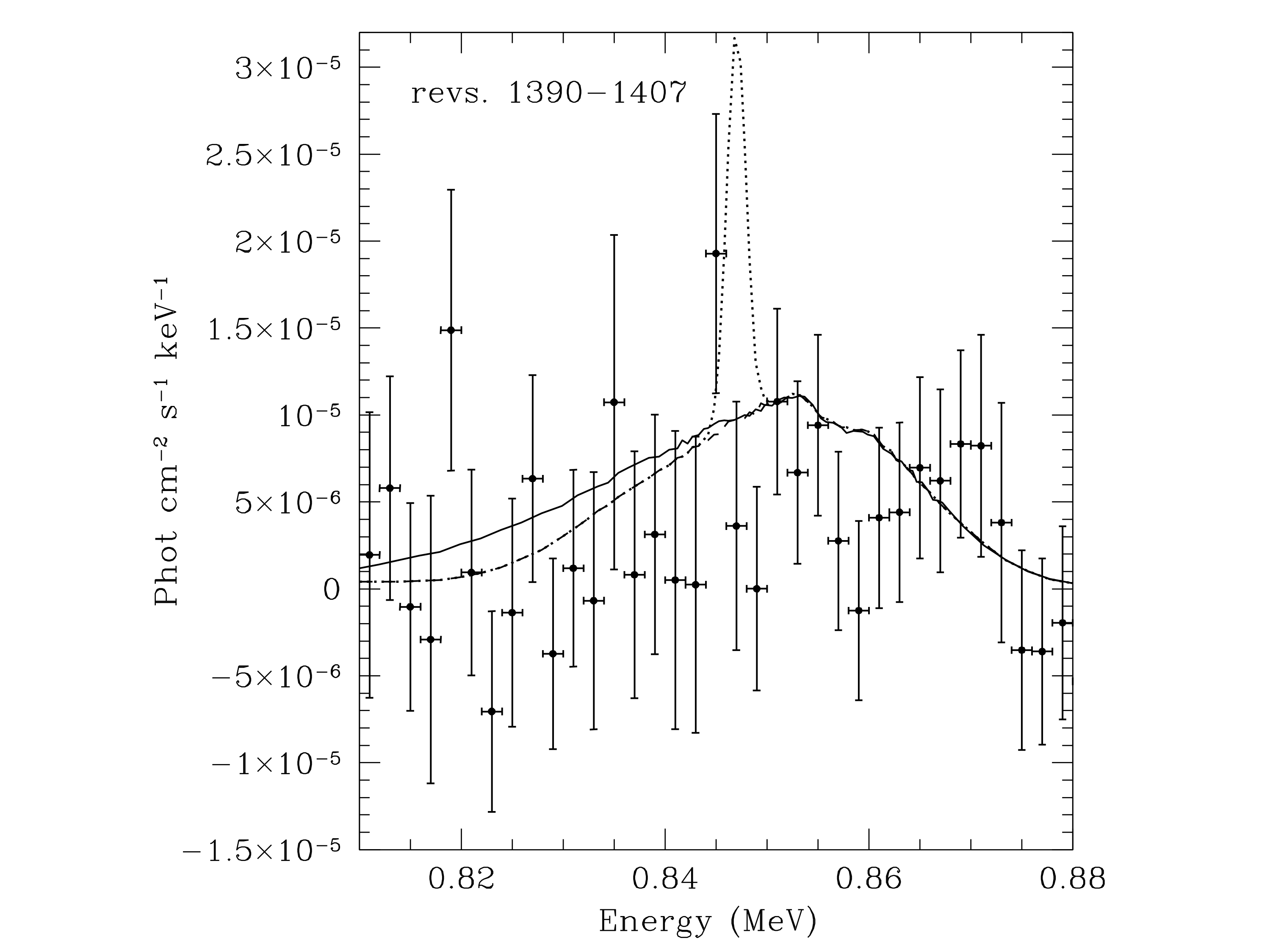}      
  \caption{  High resolution gamma-ray spectrum 50-100 days after the explosion around the 847 keV \element[ ][56]{Co} line. Bins are 2 keV wide. The dashed line represents the DDT1p4 model. The solid line represents the emission of the 3Dbball model (the DDT1p4 plus a plume of 0.07 M$_\odot$ of \element[ ][56]{Ni} out the equatorial plane). The dotted line represents the DDT1p4 model plus a plume of 0.06 M$_\odot$ of  \element[ ][56]{Ni} confined within the equatorial plane.  All of them have been computed seventy five days after the explosion. 
              }
         \label{fig13}
   \end{figure}

The presence of a plume able to produce the broad redshifted \element[ ][56]{Ni} lines claimed here  is compatible with the observed properties of the \element[ ][56]{Co} lines \citep{chur14a,chur15}  at late times. Figure~\ref{fig13} shows at high resolution (2 keV) the flux averaged over days 50 to 100 after the explosion. At this epoch the debris are almost transparent to gamma rays and the presence of the plume has a negligible effect on the spectrum since its contribution is smeared over a large energy interval as a consequence of its broadness. On the contrary, if the plume had contained almost all this amount of \element[ ][56]{Ni} in the equatorial plane and  was seen perpendicularly, it would have produced a prominent spike similar to or narrower than the one represented in the figure. Such spike would not have introduced dramatic effects at low resolution, but at the resolution provided by SPI it should be detectable. Just as an example, adding a ring of 0.06 M$_\odot$ of \element[ ][56]{Ni} in the equatorial plane able to produce a narrow (FWHM $\approx$ 2.23 keV) feature at 812 keV,  when \element[ ][56]{Ni} was detected, should produce, 75 days after the explosion, a flux at the nominal energy of the 847 keV \element[ ][56]{Co} line $\sim 4.4 \sigma$ larger than the observed value, as it can be seen in Figure~\ref{fig13}, which can be interpreted as a rejection of such a model with a probability larger than 99.9\%. The maximum mass of \element[ ][56]{Ni} that could be confined within the equatorial plane is estimated to be $\sim 0.02$ M$_\odot$ ($2 \, \sigma$ level).    

The spectral model adopted to extract the line fluxes and characteristics from these early observations is the best fitting 3Dbball model. As stated in the Appendix, this model is a combination of the DDT1p4 model, which reproduces the \element[ ][56]{Co} gamma-ray lines observed in Late Observations of SN2014J \citep{chur14a} and the optical light curve, and a \element[ ][56]{Ni} plume that accounts for the observed redshifted line.  The best fit to the SPI data (120-1350 keV band) is obtained with a mass of \element[ ][56]{Ni} of $0.077\pm 0.040$ M$_\odot$ in the plume and $0.60 \pm 0.29$ M$_\odot$ in the central body (Figure~\ref{fig5}). These best fit values were obtained with a $\chi^2 = 210.72$ for a d.o.f. of 244 (246 spectral bins and 2 free parameters: the mass of \element[ ][56]{Ni} in the plume and in the central body). Notice that this last value for the central body is in agreement with the mass derived from the optical light curve measurements (Section 2) and with the mass obtained with the analysis of the gamma-ray emission at the late epoch \citep{chur14a}.

Finally, it is worth mentioning that if this 3Dbball model is adopted as a reference, Table~\ref{tmchi} shows that DETO and DDTe models can be rejected at $\sim 7 \sigma$ and $\sim 4 \sigma$ level respectively,  but the remaining ones differ by $< 3 \sigma$ and cannot be formally rejected. Also, Figure~\ref{fig6} shows that the flux excess found at $\sim 730$ keV that was attributed to the redshifted 750 keV \element[ ][56]{Ni} line in Section 2  with a significance of 2.1 $\sigma$),  represents a flux of $(1.2 \pm 0.7) \times10^ {-4}$ ph s$^{-1}$cm$^{-2}$ ($\sim$ 1.7 $\sigma$), when the continuum underneath the line is included in the fit. Although the significance of the 730 keV line flux is not enough to claim its detection, this excess  reinforces the plausibility the \element[ ][56]{Ni} plume hypothesis.

\section{Conclusions}
SN2014J has been observed with all the instruments on board of \emph{INTEGRAL} just around the maximum of the optical light curve for a period of $\sim 10^6$ seconds. The optical light curve measured with the OMC is in agreement with the light curves obtained from the ground and can be explained by a delayed detonation model that synthesizes $\sim 0.65$ M$_\odot$ of \element[ ][56]{Ni} (see Fig.~\ref{fig1}). 

As it has previously been  stressed \citep{chur15}, despite its distance, SN2014J is a weak gamma-ray source and the results  are sensitive to various aspects of the data analysis. The main improvement with respect to previous analysis \citep{dieh14} is the clear detection of the gamma-ray signal by both instruments, SPI and ISGRI, with significances of 5$\sigma$ (see figures~\ref{fig3} and \ref{fig9}) during this period of observation at the position of SN2014J, removing the  $\sim 2^o$ offset present in the previous analysis, and confirming the idea that the light curves of SNIa are powered by the decay of \element[ ][56]{Ni}. Surprisingly,  we found in the SPI data evidences for a broad, redshifted feature that corresponds to the 158 keV emission of this isotope.  Given this energy and broadness, this feature cannot be confused with any residual instrumental line. Furthermore,   we have also found in the ISGRI data an emission excess during orbits 1380-81 and 1384-85 at a 3.8 $\sigma$ level (Table~\ref{tisgri}). These emission excesses are well above the predictions of conventional 1D models (Fig.~\ref{fig11}) and are separated by a dip that is compatible with the free decay of \element[ ][56]{Ni}, although the low significance of this dip prevents any definite consideration about the temporal variability.  A similar behavior is suggested by SPI {\bf data}, an excess of emission followed by a decline, but once more the poor significance prevents any definite conclusion.  A possible explanation of this behaviour is that during the SN2014J event, an almost $\gamma$-ray transparent plume made of $\sim 0.08$ M$_\odot$ of \element[ ][56]{Ni} was ejected with an expansion velocity of $\sim 30,000$ km/s and a dispersion velocity of  $\sim 10,000$ km/s that is globally receding from the observer with a velocity of $\sim 6,000$ km/s. The significance of this additional  \element[ ][56]{Ni}, obtained by fitting the redshifted and broadened 158 keV, 750 keV and 812 keV lines above the DDT1p4 model with SPI data, is  $\sim 3 \, \sigma$  (see Figure 4).

\citet{chur14a,chur15} reported that the fluxes and spectra  of the gamma-rays emitted by  \element[ ][56]{Co} during the maximum of the light-curve were in broad agreement with the predictions of classical spherically-symmetric theoretical models of SNIa based on either the deflagration or the delayed detonation paradigms. Figure~\ref{fig13} shows that the introduction of an extra emission caused by the existence of a radioactive plume  with the characteristics proposed here predicts a late-time spectrum that is still in accordance with the spectrum obtained during the late epoch (55-100 days after the explosion). 

In any case, the significance of the signal prevents any firm conclusion about the behavior of phenomena changing with a time scale of the order of the \element[ ][56]{Ni} decay time. Taking into account Table~\ref{tmchi} and adopting a conservative point of view, the data are largely consistent with the standard delayed detonation model \citep{chur15} without excluding the presence of small amounts of \element[ ][56]{Ni} at the surface if the lines are broad.

It is evident that if the significance of the redshift and the width of \element[ ][56]{Ni} lines found in the observations of INTEGRAL was enough, the gamma-ray behavior would introduce strong constrains on  the  acceptable  models  for  SN2014J. If confirmed in other supernovae, it could be concluded that conventional models starting with the ignition of the central regions of a C/O white  dwarf  and  keeping  the  radioactive  material  confined  in  the  innermost  layers would not be appropriate to account for the observed properties at this early epoch and that, at least in these cases, additional possibilities should be considered. Sub-Chandrasekhar models, i.e. C/O white dwarfs with a mass not necessarily near to the critical mass that explode as a consequence of the ignition of a freshly accreted He-envelope \citep{woos94}, produce \element[ ][56]{Ni} at the surface and could, in principle,  account for the observations if the mass of these layers is small enough \citep{pakm13,guil10,fink10,garc99}. Three dimensional models, like Pulsating Reverse Detonations (PRD) \citep{brav09,brav09a}, Gravitationally Confined Detonations (GCD) \citep{plew04}, and collisions of white dwarfs in double-degenerate binaries or multiple systems \citep{kush13,azna13} could also provide scenarios  with \element[ ][56]{Ni} present in  the  outer  layers, although in the last case the collision would require the presence of massive white dwarfs to achieve the observed amount of \element[ ][56]{Ni} \citep{garc13}.
Obtaining similarly extensive INTEGRAL data on additional SNIa would be of the maximum interest not only to ascertain if SN2014J is a representative event, but also to constrain the models for SN2014J like events. Nevertheless, given the existence of several SNIa subtypes, only a high sensitivity detector  would be able to provide a statistically representative sample of gamma observations. Finally, it is necessary to emphasize that the implications of the reported asymmetrical features on cosmological applications of SNIa have still to be determined.

\begin{acknowledgements}

This  work was  supported by the  MINECO-FEDER grants ESP2013-47637-P (JI), AYA2012-39362-C02-01 (AD), AYA2013-40545 (EB), ESP2013-41268-R (MH), and AYA2011-22460 (ID)  by the ESF EUROCORES Program EuroGENESIS (MINECO grant EUI2009-04170), by the grant 2009SGR315 of the Generalitat de Catalunya (JI), and by the Ministerium f\"ur Bildung und
Forschung via the DLR grant 50.OG.9503.0. NER acknowledges the support from the European Union Seventh Framework Programme (FP7/2007-2013) under grant agreement n. 267251 "Astronomy Fellowships in Italy" (AstroFIt). NER is also partially supported by the PRIN-INAF 2014 with the project "Transient Universe: unveil- ing new types of stellar explosions with PESSTO”.
The SPI project has been completed under the responsibility and leadership of CNES, France. ISGRI has been realized by CEA with the support of CNES. We acknowledge the INTEGRAL Project Scientist Erik Kuulkers (ESA, ESAC) and the ISOC for their scheduling efforts, as well as the INTEGRAL Users Group for their support in the observations.
\end{acknowledgements}

\bibliographystyle{aa}
\bibliography{sn2014J}

\appendix

\section{Theoretical models}
\label{stm}

The gamma-ray spectrum of SNIa depends on the total amount of  \element[ ][56]{Ni} and its distribution within the expanding debris, which in turn depends on the burning regime of the explosion. In the case of one dimension models, three burning modes have been identified. Table~\ref{tm} displays the main characteristics of the models used in the present study.

Pure detonations (DETO), in which carbon is ignited in the centre of a carbon-oxygen white dwarf near the Chandrasekhar’s mass and the burning propagates supersonically in such a way that the star is completely incinerated to the Fe-peak elements \citep{arne96}. This model is incompatible with all the existing observations, including those obtained by INTEGRAL in the case of SN2011fe \citep{iser13}  and SN 2014J \citep{chur14a}. It is also representative of the most massive models computed by \citet{fink10}.

Sub-Chandrasekhar detonations  (SCH) assume white dwarfs with arbitrary masses accreting helium from a companion in such a way that when the mass of this freshly accreted envelope reaches a critical value, He ignites at
 the bottom and induces the explosion of the white dwarf \citep{woos94}. The main argument against these models has so far been the non detection of significant amounts of \element[ ][56]{Ni} and \element[ ][56]{Co} moving at high velocities. The SC1F and SC3F (E.Bravo, unpublished) are SCh models equivalent to models 1 and 3 of \citet{fink10}.

Deflagration (DEF) models assume that the star is ignited in the central regions and the burning front propagates subsonically through all the star in such a way that the outer layers can expand and avoid complete incineration. The prototype is the W7 model \citep{nomo84}.

Delayed detonations (DDT) start as a deflagration in the centre and, when the flame reaches a density of few times $10^7$ g cm$^{-3}$ it turns into a detonation. Because of the low densities, characteristic burning times are too long and matter in these layers is not completely incinerated to \element[ ][56]{Ni} and only intermediate mass elements are profusely produced during this regime, in agreement with the observations \citep{khok91}.   The DDTe \citep{bade05} model is an example. 
Pulsating delayed detonations (PDD), a subtype of DDT model, assume that the burning front starts at the centre, but the flame moves so slowly that it is quenched by the expansion of the white dwarf. After reaching the maximum expansion, the star contracts and triggers the explosion \citep{khok91}. 

 \begin{figure}
   \centering
 \includegraphics[width=\hsize]{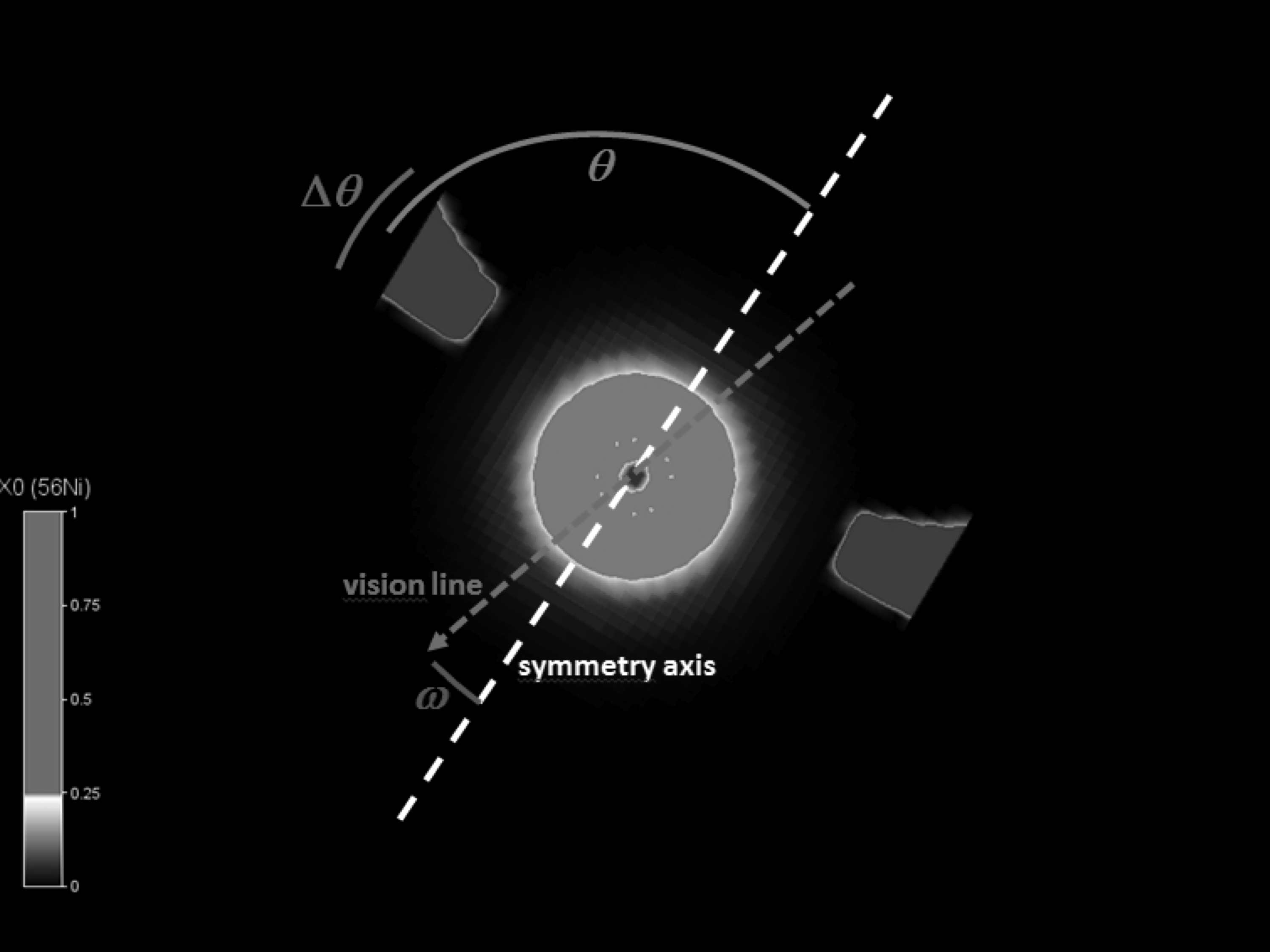}      
  \caption{ Meridional cut of a phenomenological scenario to account for the early gamma-ray emission. Two components are assumed, a central spherically symmetric remnant that contains the bulk of mass and radioactive material resulting from the explosion of the white dwarf plus a conically shaped structure made of almost pure \element[ ][56]{Ni} expanding with velocities large enough to avoid being caught by the inner material. }
         \label{figA1}
   \end{figure}

The models used here are the same as in \citet{iser13} plus the models DDT1p4 and 3Dbball. The first one was tailored to broadly reproduce the optical light curve of SN2014J. This model is centrally ignited at a density of $2\times10^9$ g cm$^{-3}$ and makes the transition deflagration/detonation at $1.4 \times 10^7$ g cm$^{-3}$. The total mass of \element[ ][56]{Ni} produced is 0.65 M$_\odot$, the mass ejected is 1.37 M$_\odot$, and the kinetic energy is $1.32 \times 10^{51}$ ergs. The 3Dbball model is essentially the DDT1p4 model plus a plume of radioactive material as depicted in Fig.~\ref{fig12}. Figure~\ref{figA1} shows  the different parameters that characterize the model. Although a full set of values was explored, a reasonable choice of parameters is: mass of  \element[ ][56]{Ni} in the conically shaped structure 0.04 - 0.08 M$_\odot$, expansion velocities $v_{\rm i}= 25,000$, $v_{\rm s} = 35,000$ km/s, while $\theta$, $\delta \theta$ and $\omega $ have to be in agreement with observed recession, $\sim  6,000$ km/s, and dispersion, $\sim 10,000$ km/s velocities as suggested by the redshifted Ni-lines. 

\begin{table}
\caption{Kinetic energy (K) and mass of \element[ ][56]{Ni} produced by different models of explosion (1 foe = $10^{51}$ erg).}             
\label{tm}      
\centering                  
\begin{tabular}{c c c}        
\hline\hline                 
Model  & $K$ (foe) &  $M_\mathrm{Ni}$  (M$_\odot$)                       \\  
 \hline                        
DETO     &    1.44   & 1.16      \\
SC3F      &    1.17   & 0.69      \\
W7         &    1.24   & 0.59      \\
DDTe     &    1.09   & 0.51      \\
SC1F      &    1.04   & 0.43      \\
DDT1p4 &    1.32   & 0.65      \\
\hline                                   
\end{tabular}
\end{table}

The gamma-ray spectrum has been obtained from a recently updated three dimensional generalization of the code described in \citet{gome98,miln04,iser08}. The initial model was obtained adding to the output of the DDT1p4 model a conical ring as described before and allowing a homologous expansion.  Given the expansion velocity that has been assumed to account for the redshift and the broadness of the line, the plume has to be clearly above the equator, $\theta \approx 78\degr$, and the angular thickness $\Delta \theta \approx 12\degr$. The line of sight has to be close to the axis of symmetry ($\omega \la  12^o$) since for larger values the 158 keV line would evolve towards a double peaked shape that does not seem consistent with the SPI spectrum. Furthermore, the lack of substantial polarization in the optical at the early epoch also favours small values of $\omega$ \citep{pata14}.

\end{document}